\documentclass[12pt]{article} 
\pdfoutput=1
\usepackage{amsmath, amsfonts, amssymb}
\usepackage{xcolor}
\usepackage[margin=0.75in]{geometry}

\usepackage{graphicx}
\usepackage{epstopdf, epsfig}
\usepackage{graphicx} % Required for inserting images
\usepackage[mmddyyyy]{datetime}
\usepackage{amsmath}
\usepackage{multicol}
\usepackage{longtable}

\usepackage{booktabs}
\usepackage{enumitem}
\usepackage{subcaption}

\usepackage{fancyhdr}
\usepackage{amsmath, amssymb}
\usepackage{algorithm}
\usepackage[round]{natbib}
\usepackage{algpseudocode}
\usepackage{hyperref}
\usepackage{array, booktabs, makecell} %makecell for breaking table header to multiple lines

\newcommand{\edit}[1]{\color{black}#1 \color{black}}

\title{Role of flow topology in wind-driven wildfire propagation}

\author{Siva Viknesh$^{1,2}$  \and  Ali Tohidi$^{3,4}$  \and Fatemeh Afghah$^{5}$  \and Rob Stoll$^{1}$  \and Amirhossein Arzani$^{1,2}$ }

\date{}

\begin{document}

\maketitle

\begin{center}
$^{1}$ Department of Mechanical Engineering, University of Utah, Salt Lake City, UT, USA \\
$^{2}$Scientific Computing \& Imaging Institute, University of Utah, Salt Lake City, UT, USA \\\edit{
$^{3}$Department of Fire Protection Engineering, University of Maryland, College Park, MD, USA\\
$^{4}$NSF-IUCRC Wildfire Interdisciplinary Research Center (WIRC), San Jos\'e, CA, USA \\}
$^{5}$Department of Electrical and Computer Engineering, Clemson University, Clemson, SC, USA
\end{center}

\bigskip

\noindent Correspondence:\\
Amirhossein Arzani,\\
University of Utah,\\
Salt Lake City, UT,  84112\\
Email: amir.arzani@sci.utah.edu
%\newpage

\thispagestyle{empty}

\begin{abstract}
Wildfires propagate through interactions between wind, fuel, and terrain, resulting in complex behaviors that challenge accurate predictions. This study investigates the interaction between wind velocity topology and wildfire dynamics, aiming to enhance our understanding of wildfire spread patterns through a simplified nonlinear convection-diffusion-reaction wildfire model, adopting a fundamental reactive flow dynamics perspective. We revisited the non-dimensionalizion of the governing combustion model by incorporating three distinct time scales. This approach revealed two new non-dimensional numbers, contrasting with the conventional non-dimensionalization that considers only a single time scale. Through scaling analysis, we analytically identified the critical determinants of transient wildfire behavior and established a state-neutral curve, indicating where initial wildfires extinguish for specific combinations of the identified non-dimensional numbers. Subsequently, a wildfire transport solver was developed, integrating upwind compact schemes and implicit-explicit Runge-Kutta methods. We explored the influence of stable and unstable manifolds in wind topology on the transport of wildfire under steady wind conditions defined using a saddle-type fixed point flow, emphasizing the role of the non-dimensional numbers. Additionally, we considered the benchmark unsteady double-gyre flow and examined the effect of unsteady wind topology on wildfire propagation, and quantified the wildfire response to varying wind oscillation frequencies and amplitudes using a transfer function approach. The results were compared to Lagrangian coherent structures (LCS) used to characterize the correspondence of manifolds with wildfire propagation. The approach of utilizing the wind flow manifolds provides valuable insights into wildfire dynamics across diverse wind scenarios, offering a potential tool for improved predictive modeling and management strategies.

\noindent\textbf{Keywords:} Wildland fire, Wind topology, Combustion, Firefront, Lagrangian coherent structures, Reactive flows

\end{abstract}

\newpage
%\linenumbers

\section{Introduction}
Wildfires have long been integral to earth's ecological processes, playing a crucial role in shaping habitat dynamics and ecosystem structures since the emergence of terrestrial vegetation. While many ecosystems rely on periodic wildfires to maintain their ecological balance, these events also pose significant threats to human infrastructure and safety, with uncontrolled and extensive wildfires often leading to catastrophic property damage, loss of life, and severe degradation of air quality. 

Wildfires represent a unique form of thermal-degradation of fuels followed by natural combustion, characterized by a self-regulating fuel supply rate intrinsically linked to the fire's behavior. Unlike other combustion processes, wildland fires rely on positive feedback mechanisms, where the intense heat generated during combustion significantly influences fire spread, with convection and thermal radiation often playing a pivotal role \citep{finneyRoleBuoyant2015}. The propagation of wildfires is generally classified into two categories: plume-driven and wind-driven fires \citep{byram1959combustion,morvan2018wildland}. Wind-driven fires are particularly hazardous due to their rapid propagation and high potential for ember casting (also known as firebrand showers), which complicates the evolution of fire spread even further due to the highly transient and intrinsically stochastic behavior \citep{tarifaFlightPaths1965,kooFirebrandsSpotting2010,Tohidi2017stochastic}. The dynamics of wildfires encompass two critical stages: ignition and the propagation of the fire front across different fuel types, particularly live fuels. Neither of these processes is fully understood or extensively explored, as various physical parameters, including the structure and condition of vegetation, topography, and atmospheric conditions such as wind patterns, air temperature, and relative humidity strongly influence them \citep{BOROUJENI2024102369,Coen17}. 

Mathematical models of wildfires provide crucial insights into the complex interactions between the firefront, vegetation, and wind patterns, which collectively constitute the fundamental mechanisms driving wildfire ignition and progression. In a comprehensive review, \citet{sullivan2009wildland} detailed the development of physics-based wildfire models, while \citet{sullivan2009wildland1} explored empirical and quasi-empirical models, with challenges and limitations in accurately simulating fire behavior discussed in \citet{sullivan2009wildland2} and \citet{or2023review}. While the simplest empirical models establish functional relationships between the fire spread rate and key parameters—such as slope, wind speed, fuel type, and bulk density—the other end of the spectrum involves the use of three-dimensional coupled computational fluid dynamics-wildfire (CFD) solvers, which aim to comprehensively capture physical and chemical interactions across a broad range of scales. Positioned between these empirical functional relations and CFD tools, several continuum physics-based wildfire models—proposed over the years by \edit{\citet{asenio, margerit2002modelling, coen2007wildland, mandel2008wildland, simeoni2011physical, burger2020exploring, vogiatzoglou2024interpretable, navas2024modeling}}—aim to describe the fundamental reactive flow physics, specifically through the nonlinear convection-diffusion-reaction (CDR) framework, while maintaining computational tractability. In this approach, the modeling involves a simplified description of combustion kinetics and thermal energy transfer, where all the physical quantities are averaged over the local plantation height, resulting in a computationally efficient system of coupled differential equations that describes the spatiotemporal evolution of the two-dimensional fields of temperature and combustibles. Furthermore, ~\citet{gollner2015towards, Verma_Ahmed_Trouvé_2022} remarked that these physics-based CDR wildfire models effectively resolve the ``flame'' scales, as the pyrolysis processes occurring at the ``vegetation'' scale are not captured by these reaction models. 
%In addition, supported by analytical methods and numerical simulations, it also elucidates the complex interactions among convection-diffusion-reaction processes and the associated mechanisms, decisively concluding that while wind speed has a profound effect on wildfire dynamics, its full implications are not yet fully understood and merits further exploration.

Numerous studies have explored the influence of wind direction on wildfire behavior within the contexts of empirical-based, physics-based CDR, and fully coupled CFD methodologies. In the realm of empirical-based wildfire models,~\citet{babak2009effect} investigated the profound impact of wind direction on the firefront's advection within a one-dimensional analytical framework, demonstrating that wind could either accelerate or decelerate fire progression depending on its flow direction while affirming the uniqueness of solutions under varying wind conditions. Similarly, ~\citet{FENDELL2001171} introduced an empirical mathematical expression that relates unidirectional wind speed to the fire spread rate for a given fuel load, based on quasi-steady firefront propagation experiments. In related work,~\citet{rossa2018empirical} developed an empirical model that accounts for wind effects, examining various fuels and their spatial orientations relative to the wind.~\citet{atchley2021effects} further advanced these studies by investigating the spatially heterogeneous distribution of fuel, while ~\citet{simpson2013large, li2021effect,bi2022experimental, bi2023experimental}, along with the references cited therein, analyzed the influence of terrain slope on wildfire dynamics.

Within the CDR wildfire framework, \citet{asenio} studied the significance of flow advection on firefronts driven by moderate, unidirectional wind over both homogeneous and heterogeneous fuel beds.~\citet{coen2007wildland} demonstrated that these physics-based wildfire models efficiently capture the temporal temperature data sampled from real-time weather measurements, thereby suggesting their applicability in short-range wildfire behavior forecasting. In their subsequent work,~\citet{mandel2008wildland} employed data assimilation methodologies to calibrate parameters within the CDR model and improve its prediction accuracy.~\citet{simeoni2011physical} studied the reliability of these models by qualitatively reproducing fire behavior observed in field experiments. ~\citet{imex} highlighted that even minor quasi-temporal variations in wind direction and firebreaks can significantly alter both the propagation direction and the spread rate of wildfires. Furthermore,~\citet{navas2024modeling} provided theoretical and numerical insights on wildfire advection under the influence of moisture content, heterogeneous fuel distribution, and non-flat topography.~\citet{reisch2024analytical} investigated the influence of diffusion and reaction parameters, along with wind speed, on wildfire propagation speed, unburned biomass fuel, and maximum achievable fire temperature by examining various simplified sub-models.

Utilizing fully coupled CFD methodologies, ~\citet{sun2009importance} investigated fire-boundary layer interactions through a fully coupled atmosphere-wildfire large eddy simulation for grassland fires. Their study highlighted how a roll-dominated atmospheric boundary layer and its intricate local flow features could influence fire spread and how the fire, in turn, affects the boundary layer.~\citet{filippi2018simulation} further corroborated the interaction between the fire and boundary layer by performing multi-scale high-fidelity large-eddy wildfire simulations, revealing the significance of intricate turbulence effects.  The studies by~\citet{clark1996coupled, coen2014simulation, coen2015high} utilize instantaneous transient wind velocity vectors  to investigate their correspondence with the advection dynamics of wildfires over realistic terrain and fuel distributions, employing the fully coupled atmosphere (weather)–wildland fire model simulations. Their findings indicate that fire progression is predominantly governed by the wind velocity vector patterns in the vicinity of the fire location.

Given the significant growing interest in understanding wind-driven wildfire transport through the wind velocity fields, the present work takes a different approach to understand how coherent flow structures in wind influence wildfire propagation patterns, rather than relying on local wind velocity vectors. This paper aims to leverage concepts from dynamical systems theory to identify and predict ``coherent patterns" in wind-driven wildfire propagation. In particular, stable and unstable manifolds, identified using Lagrangian coherent structures (LCS), have been utilized in the past to identify templates for convective transport~\citep{shadden2005definition,allshouse2015lagrangian,Haller16}. LCS has been applied across various fields, including biomedical~\citep{ArzaniShadden12}, turbulent~\citep{wilson2013identification}, environmental~\citep{bozorgmagham2013real}, and reactive flows~\citep{mahoney2012invariant}. Within the context of wildfires, recent studies from the atmospheric science community have employed the LCS framework to qualitatively predict the evolution of smoke plumes from wildfire events~\citep{curbelo2023three, allen2024smoke, jarvis2024atmospheric}. However, the direct relevance of LCS to wildfire propagation has yet to be explored. Given that wildfire transport is not purely advective, it remains unclear how flow topology can be utilized and under what conditions its role becomes significant. Gaining such fundamental insights is crucial, as it enables the prediction of wildfire spread patterns without the need to solve computationally expensive transport problems, assuming the wind velocity field is known. The attracting LCS (unstable manifolds) tend to draw transported material, such as fire lines, towards themselves, which could indicate regions of heightened risk. Conversely, the repelling LCS (stable manifolds) may potentially push fire lines away, thereby creating safer zones for positioning firefighters. 

In this paper, we consider the Asensio CDR wildfire model~\citep{asenio}, which incorporates key pathways in wildfire transport, including wind-driven convection, nonlinear diffusion due to heat and radiation, fuel-limited reaction, and simplified natural convection. The model, however, does not account for several critical factors, such as firebrands~\citep{Tohidi2017stochastic}. Additionally, we assume the wind velocity is known and provided, thereby neglecting the heating-induced buoyancy forces that can alter wind velocity—factors that have been incorporated in more complex coupled weather–wildland models~\citep{mandelCoupledAtmospherewildland2011,coen2013wrf,coen2013}. While we acknowledge the relative simplicity of the Asensio model, especially given the multiscale nature of wildfires~\citep{hadrich2021fire,hudson2020effects}, this simplicity allows us to focus on the fundamental factors and understand the role of flow topology in the interplay between key physical processes (reported using nondimensional groups). From a fundamental perspective, we approach the problem by modeling nonlinear convection-diffusion-reaction transport, where limited fuel (vegetation) acts as a source (reaction term) sustaining the high-temperature fire spread. This approach prioritizes a fundamental reactive flow mindset over a realistic, wildfire-specific modeling approach. The precise flame and combustion chemistry are not modeled within these typical CDR wildfire models, as these phenomena occur on vegetation scales much smaller than those at which wind velocity data is typically measured or modeled. Nonetheless, understanding the behavior of this particular class of reactive flows will lay the groundwork for developing more complex wildfire models in the future.

The paper is organized as follows: In Section~\ref{method_section}, we revisit the wildfire model presented by~\cite{asenio} and proceed to non-dimensionalize it, introducing previously unrecognized dimensionless numbers. This section also details the development of a wildfire transport solver based on finite difference methods. Section~\ref{scal_analy} presents a scaling analysis to identify the key factors influencing wildfire behavior within the physics-based CDR framework. In Section~\ref{wild_dyn}, we explore the characteristics of transient wildfire dynamics under steady wind conditions with saddle-type fixed points and analyze the impact of unsteady wind velocities modeled by a double-gyre flow.

\section{Methods}
\label{method_section}
\subsection{Mathematical wildfire model}
The mathematical nonlinear convection-diffusion-reaction model for wildland fire propagation, proposed by \citet{asenio}, is based on the principles of energy and species conservation, considering the quantity of fuel while disregarding its specific intricate composition. The reaction rate \( (r) \) in this model is characterized by the Arrhenius equation, \( r = A \exp{\left( -E_{\mathrm{A}}/(RT) \right)} \), and the fuel burning rate  \( (S_Y) \) is proportional to this reaction. In wildfire scenarios, sub-grid radiation flux through fuel parcels is an important mode of heat transfer, characterized by the Stefan-Boltzmann law, \( q_{\text{rad}} = 4\sigma \delta T^{3} \), \edit{adopting the local radiative model~\citep{rosseland1931astrophysik}.} Furthermore, natural convection is encapsulated by Newton's law of cooling, \( q_{\text{conv}} = h(T - T_{\infty}) \). Integrating these components--convection, radiation, reaction, and natural convection--the proposed one-way coupled model for flame propagation and fuel burning is mathematically formulated as 

\begin{subequations}
\label{dim_eq}
\begin{align}
\rho C \frac{\partial T}{\partial t} + \underbrace{ \rho C(\overrightarrow{\mathbf{v}} \cdot \nabla T)}_{\text{Convection}} &= \underbrace{\nabla \cdot \left( (4 \sigma \delta T^3 + k) \nabla T \right)}_{\text{Diffusion}} + \underbrace{s(T)^{+} \rho Y H r}_{\text{Reaction}} - \underbrace{h(T - T_{\infty})}_{\text{Natural Convection}} \\[8pt]
\frac{\partial Y}{\partial t} &= -s(T)^{+} \hspace{1mm} Y A \mathrm{e}^{-E_{\mathrm{A}} / RT} \;,
\end{align}
\end{subequations}
where $\rho$ is the density, $C$ is the constant pressure specific heat, $T$ is the absolute temperature, $\overrightarrow{\mathbf{v}}$ is the velocity vector, $\sigma$ stands for the Stefan-Boltzmann constant, $\delta$ represents the optical path length for radiation, $k$ indicates thermal conductivity, $s(T)^{+}$ characterizes the dimensional phase change function, $Y$ is the fuel mass fraction \edit{---the ratio of the mass to the initial mass,} $H$ represents the heat of combustion, $r$ is the reaction rate, $h$ is the natural heat convection coefficient, $T_{\infty}$ stands for the ambient temperature, $A$ represents the pre-exponential factor for a first order kinetic scheme, $E_{\mathrm{A}}$ is the activation energy, and $R$ is the universal gas constant. It should be noted that all physical variables in Eq.~\ref{dim_eq} are spatially averaged quantities, with the averaging performed over the fuel height, perpendicular to the ground, accounting for both air and fuel properties. Furthermore, the influence of topography can be incorporated into the wildfire model by redefining the velocity vector $\overrightarrow{\mathbf{v}}$ as proportional to the topographic gradient, as shown in~\cite{nelson2002effective},~\edit{with an additional advection correction factor as described in~\citet{grasso2020physics, nieding2024impact, navas2024modeling}.}

The mathematical model was originally non-dimensionalized using the Frank-Kamenetskii change of variables method, considering only one characteristic temporal scale (\(t_{0}\)) and spatial flame scale (\(l_{0}\)). It involves applying the following change of variables to non-dimensionalize the wildfire equation.

\begin{equation}
\begin{aligned}[t]
\xi &= \cfrac{x}{l_{0}}, & \eta &= \cfrac{y}{l_{0}}, & \tau &= \cfrac{t}{t_{0}}, & \overline{T} &= \cfrac{T-T_{\infty}}{\epsilon T_{\infty}}, & \beta &= \cfrac{Y}{Y(t_0)}, & \overrightarrow{\mathbf{w}} &= \dfrac{t_{0}}{l_{0}}  \overrightarrow{\mathbf{v}} \;,
\end{aligned}
\label{change_of_var}
\end{equation}
where \( \xi \) and \( \eta \) represent the normalized spatial lengths along the \( x \) and \( y \) directions, respectively, using the characteristic length scale \( l_{0} = \sqrt{(t_{0} k) / (\rho C)} \). \( \tau \) denotes the dimensionless evolution time normalized using the characteristic temporal scale \( t_{0} = (\mathrm{e}^{(1 / \epsilon)} \epsilon)/ (q A) \), where \( \epsilon = (R T_{\infty})/ E_{\mathrm{A}} \) represents the inverse activation energy and \( q = (H Y_0) / (C T_{\infty} ) \) is the non-dimensional reaction heat. In addition, \( \overrightarrow{\mathbf{w}} \) denotes the normalized velocity, and \( \beta \) represents the normalized fuel mass fraction. Using the non-dimensional absolute temperature (\( \overline{T} \)) and non-dimensional phase temperature (\( \overline{T}_{\mathrm{pc}} \)), the non-dimensional phase function is defined as

\begin{equation}
s(\overline{T})^{+}=
\begin{cases}
1 & \text { if } \overline{T} \ge \overline{T}_{\mathrm{pc}} \\
0 & \text { otherwise }
\end{cases}
\quad \text { with } \quad \overline{T}_{\mathrm{pc}}=\frac{T_{\mathrm{pc}}-T_{\infty}}{\epsilon T_{\infty}} \;.
\label{s_func}
\end{equation}

By using the above change of variables, the non-dimensional form of the wildfire transport model for temperature and fuel consumption fields is obtained as follows

\begin{subequations}
\label{non_dim_eq}
\begin{align}
\frac{\partial \overline{T}}{\partial \tau} + \overrightarrow{\mathbf{w}} \cdot \nabla \overline{T} &= \nabla \cdot (K(\overline{T}) \nabla \overline{T}) + f(\overline{T}, \beta) \;, \\[6pt]
\frac{\partial \beta}{\partial \tau} &= -s(\overline{T})^{+} \hspace{1mm} \frac{\epsilon}{q} \beta \mathrm{e}^{\overline{T}/(1 + \epsilon \overline{T})} \;,
\end{align}
\end{subequations}
where \( K(\overline{T})  = \overline{\kappa}(1 + \epsilon \overline{T})^{3} + 1 \) and \( f(\overline{T}, \beta)  = s(\overline{T})^{+} \beta \mathrm{e}^{\overline{T}/(1 + \epsilon \overline{T})} - \alpha \overline{T} \) are defined using the non-dimensional natural convection coefficient \( \alpha = (t_{0} h)/ (\rho C) \) and the inverse non-dimensional conductivity coefficient \( \overline{\kappa} = (4 \sigma \delta T_{\infty}^{3})/  k \). We may enforce the following zero-flux Robin boundary condition at the domain boundaries, given that the boundary is assumed to prevent wildfire propagation beyond the boundary

\begin{equation}
(\overrightarrow{\mathbf{w}}\overline{T} - K(\overline{T}) \nabla \overline{T}) \cdot \hat{n} = 0 \;.
\label{bc}
\end{equation}

Finally, the stability of the numerical methods used to solve the wildfire equation can be assessed by observing the highest temperature $(\overline{T}_{\max })$ over time. For highly accurate numerical schemes, this peak temperature should closely match, but not exceed, the maximum temperature predicted by the zero-dimensional non-diffusive combustion model at the initialized time instant, as expressed by

\begin{equation}
\overline{T}_{\max }=\frac{q}{\epsilon}\left(\beta +\frac{\epsilon}{q} \hspace{1mm} \overline{T}\right) \;.
\label{Tmax}
\end{equation}

\subsection{New non-dimensional wildfire model}
In the wildfire model described by Eq. \ref{dim_eq}, three distinct characteristic temporal scales are evident: fuel reaction ($t_r$), flow convection ($t_f$), and diffusion ($t_d$). However, \citet{asenio} assumed these temporal scales to be identical when using the Frank-Kamenetskii method to derive the non-dimensionalized form of the wildfire equation shown in Eq. \ref{non_dim_eq}. Here, we define the three distinct characteristic temporal scales for the three different processes according to Frank-Kamenetskii theory, along with the wind velocity:

\begin{equation}
t_d = \frac{\rho C l_0^2}{k}, \quad t_r = t_0 = \frac{\epsilon e^{(1/\epsilon)}}{qA}, \quad t_f = \frac{l_0}{U_\infty}, \quad \overrightarrow{\mathbf{w}}=\frac{\overrightarrow{\mathbf{v}}}{U_\infty} \;.
\label{temporal_scales}
\end{equation}

By incorporating the three temporal scales, the velocity scale normalized using $U_\infty$ representing the freestream velocity, and the change of variables outlined in Eq. \ref{change_of_var}, we derive the new non-dimensional form of the wildfire equation as follows

\begin{subequations}
\begin{align}
\frac{\partial \overline{T}}{\partial \tau}+ \frac{\overrightarrow{\mathbf{w}}}{\Phi} \cdot \nabla \overline{T} &= \frac{1}{Da} \nabla \cdot (K(\overline{T}) \nabla \overline{T}) + f(\overline{T}, \beta) \;, \\[6pt]
\frac{\partial \beta}{\partial \tau} &= -s(\overline{T})^{+} \hspace{1mm}\frac{\epsilon}{q} \beta \mathrm{e}^{\overline{T}/(1 + \epsilon \overline{T})} \;.
\end{align}
\label{new_non_dim_eq}
\end{subequations}

In this context, the Damk\"{o}hler number $\textit{Da} = t_d /t_r = (\rho C l_0^2qA)/(\epsilon e^{(1/\epsilon)}k)$ represents the ratio of the diffusion time scale to the reaction time scale, indicating how fast the reaction rate is in comparison to the diffusion rate. Similarly, Peclet number $\textit{Pe} = t_d /t_f = (\rho C U_\infty l_0)/ k$ signifies the ratio of the diffusion time scale to the flow convection time scale, elucidating the relationship between convection and diffusion. Furthermore, a new non-dimensional number, \( \Phi = t_f/ t_r = \textit{Da} /\textit{Pe} \),
%= (qAl_0)/(\epsilon e^{(1/\epsilon)} U_\infty) \)
has been identified and defined as the ratio of flow convection rate to reaction rate, thus establishing a relationship between the Damk\"{o}hler number and the Peclet number. This process of non-dimensionalization has unveiled a previously unnoticed new non-dimensional number ($\Phi$) and Damk\"{o}hler number inherent within the wildfire combustion model under consideration. It is imperative to underscore that the resulting new non-dimensional combustion model (Eq. \ref{new_non_dim_eq}) reduces to the conventional form represented by Eq. \ref{non_dim_eq} in instances where both \textit{Da} and $\Phi$ values converge to 1.

Based on reported data in the literature~\citep{FENDELL2001171, asenio, cheney2012predicting, amini2019pyrolysis, imex, Verma_Ahmed_Trouvé_2022}, the following variables govern the two identified non-dimensional numbers: the ambient wind speed \( U_\infty \), ranging from \( 10^{-3} \,\text{m} \,\text{s}^{-1} \) (no wind) to \( 25 \, \text{m} \,\text{s}^{-1} \); the flame length scale \( l_0 \), which varies from \( \mathcal{O}(10^{-1}) \) to \( \mathcal{O}(10^{2}) \, \text{m} \); and the pre-exponential factor \( A \), spanning \( 10^{6} \) to \( 10^{14} \, \text{s}^{-1} \), with higher values observed for dead fuels. These ranges were used to guide the selection of the non-dimensional numbers within our non-dimensional equation. It is important to note a limitation within the physics-based CDR wildfire model, where the variation of non-dimensional numbers are ``shaped'' by other spatially averaged physical variables accounting for both fuel and air properties ($\rho = 100\,\text{kg}\,\text{m}^{-3}$, $C = 1\,\text{kJ}\,\text{kg}^{-1}\,\text{K}^{-1}$, $k = 1\,\text{W}\,\text{m}^{-1}\,\text{K}^{-1}$, $T_\infty = 300\,\text{K}$). As pointed out by \cite{vogiatzoglou2024interpretable}, the spatial averaging process over plantation height and the refinement of the choice of mean values would constitute an independent fluid mechanics problem, which has been left for future studies aimed at improving wildfire CDR models.

For the same choice of mean values, the \textit{Da} number is found to vary from \( \mathcal{O}(1) \) to \( \mathcal{O}(10^{7}) \) considering extreme values of $l_0$ and $A$ values. On the other hand, the \( \Phi \) value, representing the convection term influence, ranges from \( \mathcal{O}(10^2) \) to \( \mathcal{O}(10^{-7}) \), computed based on the chosen range of $l_0$, $A$, and $U_\infty$ values. In this study, the newly derived non-dimensional wildfire equation is employed to investigate fire propagation under various wind conditions, with \(\textit{Da}\) ranging from \(1\) to \(10^5\), characterizing the strength of diffusion, and the wind topologies associated with \( \Phi \) values ranging from \(10^2\) to \(10^{-3}\), covering the wind regime from very calm to moderate. Furthermore, the effect of the reaction process is investigated by varying the parameter \( \epsilon \) between \(0.03\) and \(0.045\), corresponding to different activation energy values. It is anticipated that in scenarios with significantly stronger winds, the influence of convection and wind topology on wildfire propagation will become more pronounced.

%\begin{figure}  
%   \centering
%    \includegraphics[width=\textwidth,height=\textheight,keepaspectratio]{Fig_number_range.eps}
%    \caption{\edit{Influence of the pre-exponential factor \( A \) on the (a) characteristic flame length scale \( l_0 \) in meters, (b) Damköhler number \textit{Da}, and (c) the new non-dimensional number \( \Phi \), computed for \( T_\infty = 300 \, \text{K} \) and \( \epsilon = 0.03 \). The horizontal dotted lines with arrows in (b) and (c) indicate the direction along which the corresponding non-dimensional numbers are considered in the simulations performed in this study.}}
%    \label{fig:range}
%\end{figure}

\subsection{Numerical methods}
In this section, we present our approach to solving the physics-based CDR wildfire model using the finite difference method (FDM), with the additional advantage of leveraging CUDA support to facilitate simulations on GPUs. We implement an upwind compact scheme for the advective term to maintain its upwinding properties and employ central difference schemes for the diffusion term to accurately mirror its isotropic nature. Given the inherent numerical stiffness of the wildfire equation, we utilize the Implicit-Explicit Runge-Kutta (IMEX-RK) method for temporal integration, carefully distinguishing between stiff and non-stiff terms within the newly non-dimensionalized wildfire equation. Additionally, we incorporate a localized artificial diffusivity model to mitigate numerical instabilities, which is discussed below. \edit{The detailed FDM solver algorithm and the corresponding validation of the implementation are provided in Appendix~\ref{appendix:solver_alog} and Appendix~\ref{appendix:validation}, respectively. Additionally, Appendix~\ref{appendix:method_comp} presents a comparative analysis of three wildfire modeling strategies—a semi-empirical approach, the newly derived CDR model, and a fully coupled CFD framework—focusing on their rate of spread prediction capabilities.}

\subsubsection{Upwind compact FDM scheme}
The compact schemes with spectral-like resolution were initially introduced by \citet{lele} within the FDM framework. While this scheme exhibits superior performance for periodic boundary conditions due to its implicit nature, it tends to manifest numerical instabilities such as anti-diffusion and anti-dissipation near the domain boundaries when applied to non-periodic problems. Furthermore, optimization of the compact scheme for fluid flow problems was conducted by~\citet{sengupta2003analysis} through global spectral analysis, where the near-boundary numerical instabilities for non-periodic problems were mitigated by the development of higher-order explicit upwind boundary closures, effectively resolving length scales without the need for numerical filtering or damping. \edit{As proposed in~\citet{sengupta2003analysis},} the optimized \edit{fifth-order} upwind compact scheme (OUCS2) is utilized for computing the first derivative advection term at interior grid points is expressed as follows
\begin{equation}
\begin{aligned}
p_{i-1} T_{i-1}^{\prime}+T_{i}^{\prime}+p_{i+1} T_{i+1}^{\prime} &= \frac{1}{h_x} \sum_{m=-2}^{2} s_{m} T_{i+m}  \;,
\end{aligned}
\end{equation}
where $T^\prime_i$ represents the spatial derivative calculated using the information of $T_i$ at a given $i^{th}$ node for the grid spacing of $h_x$. The other relevant coefficient values are given by
\[
\begin{aligned}
p_{i \pm 1} &= D \pm \frac{\widehat{\eta}}{60}, &\quad s_{\pm 2} &= \pm \frac{F}{4} + \frac{\widehat{\eta}}{300}, &\quad s_{\pm 1} &= \pm \frac{E}{2} + \frac{\widehat{\eta}}{30}, &\quad s_{0} &= -\frac{11 \widehat{\eta}}{150}, \\
D &= 0.3793894912, &\quad F &= 1.57557379, &\quad E &= 0.183205192, &\quad \widehat{\eta} &= -2.
\end{aligned}
\]

% The following equations give the explicit boundary closures, 

% For a left boundary, \( i = 1 \) and \( 2 \):
% \[
% \begin{aligned}
% T_{1}^{\prime} &= \frac{1}{2h_x}\left(-3T_{1} + 4T_{2} - T_{3}\right) \\
% T_{1}^{\prime \prime} &= \frac{1}{h_x^{2}}\left(T_{1} - 2T_{2} + T_{3}\right) \\
% T_{2}^{\prime} &= \frac{1}{h_x}\left(\left(\frac{2\gamma}{3} - \frac{1}{3}\right)T_{1} - \left(\frac{8\gamma}{3} + \frac{1}{2}\right)T_{2} + (4\gamma + 1)T_{3}\right. \\
% &\quad \left. - \left(\frac{8\gamma}{3} + \frac{1}{6}\right)T_{4} + \frac{2\gamma}{3}T_{5}\right) \\
% T_{2}^{\prime \prime} &= \frac{1}{h_x^{2}}\left(T_{1} - 2T_{2} + T_{3}\right)
% \end{aligned}
% \]

% For a right boundary, \( i = (N-1) \) and \( N \):
% \[
% \begin{aligned}
% T_{N-1}^{\prime} &= -\frac{1}{h_x}\left(\left(\frac{2\gamma}{3} - \frac{1}{3}\right)T_{N} - \left(\frac{8\gamma}{3} + \frac{1}{2}\right)T_{N-1} + (4\gamma + 1)T_{N-2}\right. \\
% &\quad \left. - \left(\frac{8\gamma}{3} + \frac{1}{6}\right)T_{N-3} + \frac{2\gamma}{3}T_{N-4}\right) \\
% T_{N-1}^{\prime \prime} &= \frac{1}{h_x^{2}}\left(T_{N} - 2T_{N-1} + T_{N-2}\right) \\
% T_{N}^{\prime} &= \frac{1}{2h_x}\left(3T_{N} - 4T_{N-1} + T_{N-2}\right) \\
% T_{N}^{\prime \prime} &= \frac{1}{h_x^{2}}\left(T_{N} - 2T_{N-1} + T_{N-2}\right)
% \end{aligned}
% \]

% where \( \gamma = -0.025 \) for \( i = 2 \) and \( \gamma = 0.09 \) for \( i = (N-1) \).

\subsubsection{IMEX-RK method}
The fuel reaction term in the wildfire equation imposes a more severe time-step restriction than the diffusion term, which makes the long-time simulation laborious when one uses a fully explicit time integration method, and it becomes intractable for a fully implicit time integration scheme where one has to solve a system of nonlinear equations. On the other hand, the splitting technique may be helpful for solving the wildfire transport equation with the dynamics at different time scales, provided spatio-temporal balancing is ensured over time. The IMEX-RK schemes provide a viable strategy to integrate stiff and non-stiff terms of an equation, \textit{simultaneously}, using an implicit and explicit temporal scheme, respectively. In this work, we employed the robust L-stable IMEX-RK3 scheme outlined by~\citet{rkscheme} for temporal integration across all wildfire simulations investigated. Notably, \citet{imex} introduced the IMEX-RK temporal scheme for the wildfire equation while identifying stiff and non-stiff terms for the first time, and we extended their algorithm to accommodate the new non-dimensional form of the equation. We highlight  the stiff (\(\overline{T}\)) and non-stiff non-dimensional temperature (\(\widehat{T}\)) components within our newly derived non-dimensional wildfire equation,  given as follows: 

\begin{subequations}
\begin{align}
\frac{\partial \overline{T}}{\partial \tau} &= \frac{1}{Da} \nabla\cdot(K(\widehat{T}) \nabla \overline{T})+s(\widehat{T})^{+} \beta \mathrm{e}^{\widehat{T} /(1+\epsilon \widehat{T})}-\alpha \overline{T} - \frac{\overrightarrow{\mathbf{w}}}{\Phi} \cdot \nabla \widehat{T} \;, \\[6pt]
\frac{\partial \beta}{\partial \tau} &= -s(\widehat{T})^{+} \frac{\epsilon}{q} \beta \mathrm{e}^{\widehat{T} /(1+\epsilon \widehat{T})} \;.
\end{align}
\label{stiff_non_stiff}
\end{subequations}

\subsubsection{Localized artificial diffusivity model}
% The $C_\mu$ parameter is chosen in such a way that it absorbs other coefficients present in the proposed LAD model
In wildfire transport, the propagation of a firefront often manifests as a discontinuous wave, leading to the excitation of high-frequency numerical instabilities due to Gibbs' phenomenon in the vicinity of the firefront across the given fuel field. This poses a challenge for obtaining a tractable numerical solution, necessitating the use of artificial diffusion to enable very long-time integration, particularly in situations with very \textit{low} $\Phi$ values--indicative of high wind velocities. In resolving the advecting firefront, it is crucial to maintain high accuracy in regions where the solution is smooth while ensuring stability at discontinuities. To address this, we employ a localized artificial diffusion approach to attenuate high-frequency modes near the resolution limit without compromising lower-frequency ones. This method introduces artificial diffusion that varies spatially, targeting sharp discontinuities identified through higher-order spatial derivative values. Leveraging the localized artificial diffusion (LAD) model proposed by \citet{aslani2018localized}, we delineate the firefront and smooth regions employing the Heaviside function $(\mathbf{H_F})$ applied to the absolute fourth derivative of the temperature field, resulting in a calculation of localized artificial diffusion coefficient ($\mu$). Given that temperature is a scalar field, we observed that further application of Gaussian filtration is unnecessary and does not impact accuracy, thus saving computational time. The localized artificial diffusivity model used in the present work is given as follows

\begin{equation}
\Delta^2_T = \mu \nabla^2 \overline{T} \quad \;,  \quad \mu = C_\mu \mathbf{H_F}(|\nabla^4 \overline{T}|) \;.
\end{equation}

\subsection{Lagrangian coherent structures}

Lagrangian coherent structures (LCS) are pivotal in the study of complex dynamical systems~\citep{haller2000lagrangian}, particularly in fluid dynamics, where they serve as organizing templates for unsteady flows and govern the transport and mixing of passive particles over time. The simplest numerical approach for detecting LCS is through the finite-time Lyapunov exponent (FTLE) fields~\citep{shadden2011lagrangian}, which quantify the rate of separation of infinitesimally close trajectories over a finite time interval. The ridges of the FTLE field (locally maximum FTLE in the transverse direction) delineate the LCS and partition regions within the flow field that exhibit distinct flow behavior. 

Given a velocity field $\mathbf{u}(\mathbf{x}, t)$ describing the flow, the trajectory $\mathbf{x}(t)$ of a particle starting at position $\mathbf{x}_0$ at time $\widehat{t}_0$ is governed by the ordinary differential equation
\begin{equation}
\frac{d\mathbf{x}}{dt} = \mathbf{u}(\mathbf{x}, t) \;.
\end{equation}

To quantify the separation between nearby trajectories, the Cauchy-Green strain tensor $\widehat{\mathbf{C}}_{\widehat{t}_0}^{t_1}(\mathbf{x}_0)$ is computed as:
\begin{equation}
\widehat{\mathbf{C}}_{\widehat{t}_0}^{t_1}(\mathbf{x}_0) = \left(\frac{\partial \mathbf{x}(t)}{\partial \mathbf{x}_0}\right)^\top \frac{\partial \mathbf{x}(t)}{\partial \mathbf{x}_0} \;,
\end{equation}
where $\frac{\partial \mathbf{x}(t)}{\partial \mathbf{x}_0}$ represents the flow map gradient, describing how initial positions evolve over time. The largest eigenvalue $\overline{\lambda}_{\text{max}}$ of the Cauchy-Green tensor defines the FTLE as
\begin{equation}
\widehat{\sigma}_{\widehat{t}_0}^{t_1}(\mathbf{x}_0) = \frac{1}{|t_1 - \widehat{t}_0|} \ln \sqrt{\overline{\lambda}_{\text{max}}} \;.
\end{equation}

The ridges of the FTLE field, \(\widehat{\sigma}_{\widehat{t}_0}^{t_1}(\mathbf{x}_0)\), signify the presence of LCS, delineating boundaries between regions exhibiting distinct Lagrangian behavior. In the present work, where we consider the 2D unsteady double-gyre, the FTLE fields were computed using the TBarrier package \citep{encinas2023tbarrier}. It is important to acknowledge that the LCS is based on passive transport by the velocity field, which does not fully apply to wildfires. Our objective here is to explore the extent to which LCS can influence firefront propagation. Future research should focus on developing generalized LCS frameworks~\citep{balasuriya2018generalized} to improve the correspondence with wildfires.

%In the context of wildfire dynamics, LCS is instrumental in predicting the movement of firefronts under varying wind conditions. \citet{shadden2011lagrangian} demonstrated that instantaneous streamlines, derived from instantaneous wind velocity data, often fail to capture these material transport barriers over time, leading to significant misinterpretations. When analyzing wildfire propagation, particularly under transient wind conditions, instantaneous LCS is necessary, to understand how different regions within the flow field influence transient fire spread. By studying LCS, one can utilize the repelling or attracting coherent structures in the flow to predict areas that may either accelerate or inhibit fire spread. Furthermore, for the present investigation, FTLE fields are computed using the TBarrier package \citet{encinas2023tbarrier}, allowing instantaneous calculation of LCS at a given $\tau$.

\section{Scaling analysis}
\label{scal_analy}
\begin{figure}[h!]  
    \centering
    \includegraphics[scale=0.7,keepaspectratio]{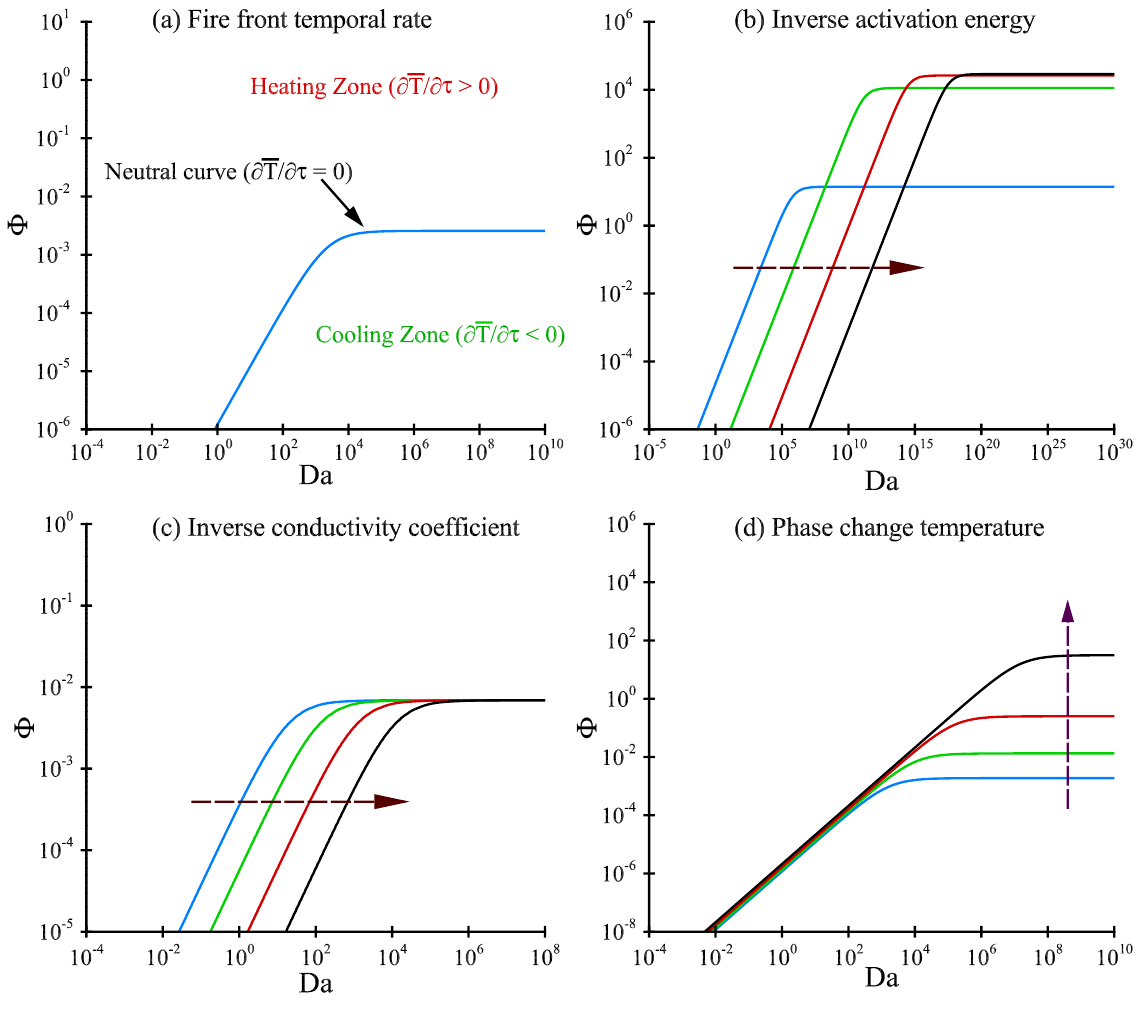}
    \caption{Temporal evolution of the initialized firefront temperature under the influence of various parameters. (a) A zero-valued contour of $\partial \overline{T}/\partial \tau$ is highlighted, representing the neutral curve along with the heating ($\partial \overline{T}/\partial \tau > 0$) and cooling ($\partial \overline{T}/\partial \tau < 0$) zones over the \textit{Da}--$\Phi$ non-dimensional plane. (b) Inverse activation energy $\epsilon$, (c) inverse non-dimensional conductivity coefficient $\overline{\kappa}$, and (d) non-dimensional phase change temperature $\overline{T}{\text{pc}}$ are shown in the same plane. The dotted arrows indicate the incremented directions of these dimensionless parameters.}
    \label{fig:scaling}
\end{figure}

%Drawing upon these insights, it becomes feasible to select appropriate values for these dimensionless parameters, thus facilitating the execution of relevant wildfire numerical simulations

First, we perform a scaling analysis to understand the influence of the dimensionless parameters on the underlying transient reactive flow dynamics within the newly derived non-dimensional physics-based CDR wildfire equation. Below, we introduce characteristic scales that, under the specified assumptions, transform the spatial derivative terms in the differential equations into an algebraic form. This analytical process provides valuable insights into the relative significance of various terms and parameters, thereby unveiling the critical determinants governing transient wildfire behavior within the physics-based CDR model. The chosen characteristic scales for scaling analysis are as follows

\begin{equation}
\overline{T} \sim \overline{T}_{\text{max}} \;, \quad \nabla \overline{T} \sim \frac{(\overline{T}_{\text{max}} - \overline{T}_{\text{pc}})}{\overline{h}_x} = \overline{T}_\Delta \;, \quad \beta \sim \beta_{\text{max}}\;, \quad |\overrightarrow{\mathbf{w}}| \sim U_\infty \;.
\label{eq:scaling_scales}
\end{equation}
Utilizing the established characteristic scales for temperature and fuel, we proceed with the following assumptions for the scaling analysis: $(i)$ Uniform wind velocity is assumed using the \(U_\infty\) scale, disregarding spatial velocity gradients. $(ii)$ The firefront consistently attains \(\overline{T}_{\text{max}}\) regardless of fuel spatial distribution. $(iii)$ A maximum value of fuel distribution (\(\beta_{\text{max}}\)) is assumed everywhere, ignoring heterogeneous distribution across space. $(iv)$ Assuming that the firefront thickness (\(\overline{h}_x\)) is temporally invariant, the temperature gradient is approximated as the difference between \(\overline{T}_{\text{max}}\) and \(\overline{T}_{\text{pc}}\) over the firefront thickness. Moreover, \(\overline{T}_{\text{max}}\) can be calculated using the analytical formulation given in Eq. \ref{Tmax}. Under these conditions, we derive the algebraic form of Eq. \ref{new_non_dim_eq}, representing the temporal evolution of initialized firefront temperature, as given below

\begin{equation}
\begin{aligned}
\frac{\partial \overline{T}}{\partial \tau} &\sim \underbrace{\frac{1}{Da} \left\{ \overline{T}_\Delta^2 \left[ 3 \overline{\kappa} \epsilon (1+\epsilon \overline{T}_{\text{max}})^2 \right] 
+ K \frac{\overline{T}_\Delta}{\overline{h}_x} \right\} + \beta_{\text{max}} \mathrm{e}^{\overline{T}_{\text{max}}}}_{\text{Heating term}} - \underbrace{\left[ \frac{U_\infty}{\Phi} \overline{T}_\Delta + \alpha \overline{T}_{\text{max}} \right]}_{\text{Cooling term}} \;.
\end{aligned}
\label{eq:new_non_dim_eq_scal}
\end{equation}

Based on the temperature temporal rate expressed in Eq. \ref{eq:new_non_dim_eq_scal}, it is observed that diffusion and reaction processes provide the energy required to heat the initialized firefront, while natural and forced convection processes remove heat energy, cooling the firefront. Interestingly, for certain combinations of the \textit{Da} and $\Phi$ numbers, along with other given dimensionless parameters, these heating and cooling processes balance each other. At this balance point, the fire neither grows nor decays from its initialized strength, a state referred to as the neutral curve, as shown in Fig. \ref{fig:scaling}(a). In other words, the neutral curve represents a state where the temperature's temporal rate, $\partial \overline{T}/\partial \tau$, is zero, demarcating the zones on the \textit{Da}--$\Phi$ plane where heating ($\partial \overline{T}/\partial \tau > 0$) and cooling ($\partial \overline{T}/\partial \tau < 0$) occur. Notably, when depicted on a log-log scale, the neutral curve exhibits a nearly linear growth, particularly saturating at higher \textit{Da} values. Furthermore, a power law fitting analysis was conducted, expressing $\Phi = m Da^n$, to quantitatively characterize this growth behavior. This mathematical regression yielded an exponent value of $n = 0.8564$ (slope in log-log scale) and an amplitude of $m = 5.882\times10^{-5}$ (offset in log-log scale), providing valuable insights into the intricate relationship between \textit{Da} and $\Phi$ regarding the temperature temporal rate of change. Moreover, the influence of other significant dimensionless parameters on the neutral curve is outlined in Fig. \ref{fig:scaling}. It is observed that an increase in the inverse activation energy notably extends the heating zone in comparison to the cooling zone, particularly evident at higher values, as depicted in Fig. \ref{fig:scaling}b. Similarly, Fig. \ref{fig:scaling}c demonstrates that increments in the inverse dimensionless thermal conductivity expand the heating zone by shifting the linear curve, while the saturation line remains unaffected. Lastly, as illustrated in Fig. \ref{fig:scaling}d, with an increase in the phase change temperature, the linear curve remains unaltered while the saturation line shifts upward, resulting in an expanded cooling zone. Interestingly, these influential parameters alter the neutral curve \textit{only} by modifying its amplitude (or offset) without changing the exponent (or slope) value, thereby resulting in the exponent (slope) value remaining \textit{invariant} in the considered physics-based CDR wildfire model.

\section{Wildfire Dynamics}
\label{wild_dyn}
\subsection{Steady wind velocity: A saddle-type fixed point}

\begin{figure}[h!]  
    \centering
    \includegraphics[width=.95\textwidth]{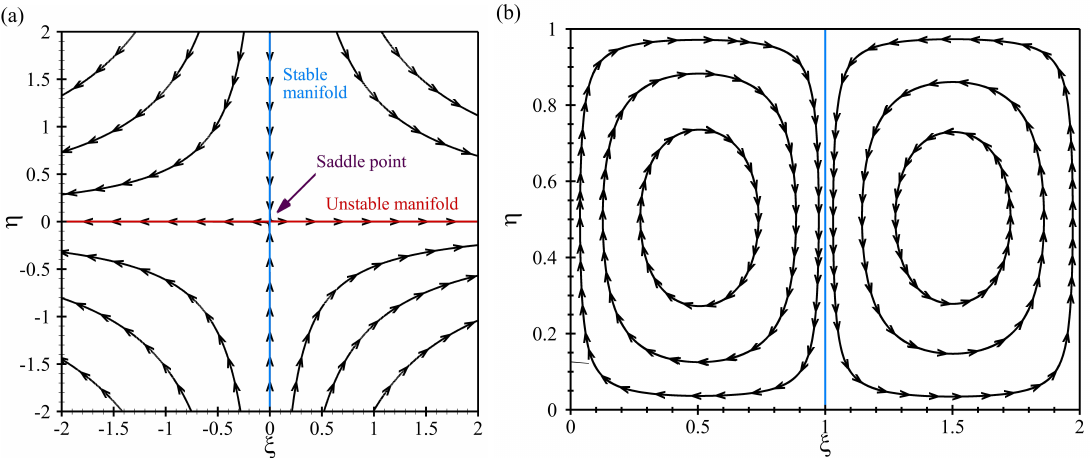}
    \caption{The two wind patterns considered in this study. (a)  Steady wind topology showcasing both stable (blue) and unstable (red) manifolds, with a saddle point positioned at the origin, superimposed with velocity streamlines. (b) Unsteady double gyre wind velocity at $\tau = 0$, where the dividing streamline (blue) shows the boundary between the two vortices. The vertical dividing streamline undergoes translation motion along the $\xi$ axis, oscillating about the value of $1$, with a given amplitude and frequency over time.}
    \label{fig:wind_velocity}
\end{figure}

We investigate the reactive flow dynamics of the simplified physics-based CDR wildfire model within the context of steady wind velocity fields. By utilizing a saddle-type fixed point in the wind velocity topology, characterized by both attraction and repulsion, we analyze the complex dynamics of fire spread, examining how wildfires respond to varying wind patterns across the spatial domain and providing valuable insights into their behavior under diverse conditions. The selected saddle-type wind topology, with stable and unstable eigenvalues of \(-1\) and \(1\), is constructed by formulating wind velocity components \( u = \xi \) and \( v = -\eta \), ensuring the conservation of mass. As depicted in Fig. \ref{fig:wind_velocity}a, the saddle point is located at the origin \((0, 0)\). The stable (repelling) and unstable (attracting) manifolds are also shown. Consistent parameters are used across all simulations to ensure coherence and comparability, including a square domain length of \( 4 \), spanning coordinates from \(-2\) to \(2\) with $\beta = 1$ everywhere across a \(256 \times 256\) grid. A fixed time step of \(\Delta \tau = 10^{-7}\) is used for time integration, and additional parameters include \( q = 1 \), \( \alpha = 10^{-3} \), \(\overline{\kappa} = 0.1\), and \( \overline{T}_{\text{pc}} = 3 \). The initial Heaviside square firefront temperature is uniformly set to \(\overline{T} = 31\), positioned at coordinates \((-1.05, 1.05)\) with a square side length of \(0.1\), and a normalized maximum wind speed of \(|\overrightarrow{\mathbf{w}}| = 1\). The corresponding dimensional initial fire temperatures are $T_{\text{pc}} = 327$ K (\( \overline{T}_{\text{pc}} = 3 \)) and $T = 527$ K (\(\overline{T} = 31\)), for $\epsilon = 0.03$ and ambient temperature $T_\infty = 300$ K, which are sufficient to initiate the flammable exothermic process, as noted by~\citet{ seron2005evolution, sudhakar2011experimental, vogiatzoglou2024interpretable}. To maintain consistency across simulations, a stabilization parameter of \( C\mu = 0.80\) is used, although this LAD model may not be essential for scenarios with moderate \(\Phi\) and \( Da \) values. To characterize the transient behavior of the initialized firefront, we track the instantaneous locations of four firefronts over time: the top firefront (\(F_T^Y\)) advecting along the positive \(\eta\) direction, the bottom firefront (\(F_B^Y\)) along the negative \(\eta\) direction, the right firefront (\(F_R^X\)) along the positive \(\xi\) direction, and the left firefront (\(F_L^X\)) along the negative \(\xi\) direction. The corresponding group velocities are indicated by $V_T^Y$, $V_B^Y$, $V_R^X$, and $V_L^Y$, respectively. This tracking method provides a detailed understanding of the firefronts' dynamics and progression under varying convection-diffusion-reaction processes.

\subsubsection{Role of Diffusion--\textit{Da} number}
\begin{figure}[h!]   
    \centering
    \includegraphics[width=1\textwidth]{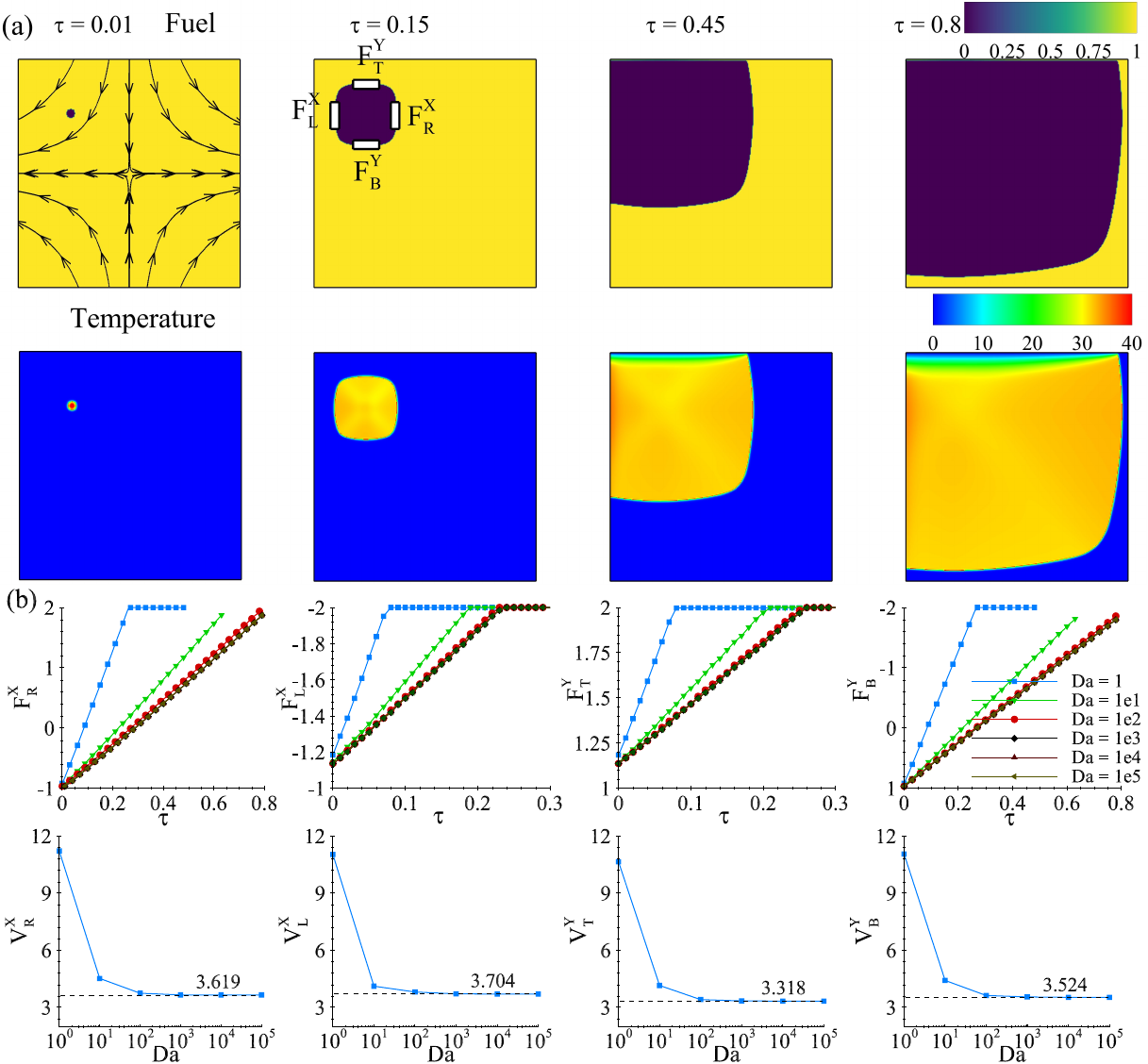}    
    \caption{Spatio-temporal evolution of the Heaviside firefront under a saddle-type steady wind velocity field over a uniformly distributed fuel bed, computed across a range of \textit{Da} values \( \in (1, 10, 10^2, 10^3,  10^4,  10^5)\), with \( \Phi = 1.0 \) and \( \epsilon = 0.03 \). (a) Fuel and temperature fields at various time instants are displayed in the top and bottom rows, respectively, for \(Da = 10^3\), with streamlines superimposed at \( \tau = 0.01 \). At \( \tau = 0.15 \), the firefront locations are marked by white strips, labeled as the top firefront (\( F_T^{Y} \)), bottom firefront (\( F_B^{Y} \)), right firefront (\( F_R^{X} \)), and left firefront (\( F_L^{X} \)), with superscripts indicating their movement direction. (b) The instantaneous spatial advection of the firefronts and their corresponding time-averaged group velocities are shown, depicted in the top and bottom rows, respectively. The dotted horizontal line represents the asymptotically converged group velocity of the firefronts at an infinite \( Da \) number.}

    \label{fig:da_number}
\end{figure}

First, we focus on understanding the role of diffusion on the transient behavior of wildfires by varying the \( Da \) number across a range of values \(  (1, 10, 10^2, 10^3,  10^4,  10^5)\), while fixing \(\Phi = 1.0\) and \(\epsilon = 0.03\) constant. The simulation results for \( Da = 10^3 \) are presented in Fig. \ref{fig:da_number}a, providing insights into the spatio-temporal evolution of the firefront over the uniform fuel bed. At \(\tau = 0.15\), the positions of the four tracked firefronts are visually represented by white strips overlaying the fuel bed, progressively advancing along their respective directions over time, while the provided fuel bed undergoes complete gradual consumption, contributing to the fire's progression without being influenced by the manifolds. In the top row of Fig.~\ref{fig:da_number}b, the instantaneous positions of these firefronts are depicted for the chosen distinct \textit{Da} values, indicating the group velocity through their slopes. The group velocity is calculated instantaneously for the given \textit{Da} and firefronts, followed by temporal averaging. The corresponding temporally averaged group velocities of the firefronts are illustrated in the bottom row of Fig. \ref{fig:da_number}b for different \textit{Da} values. 

Interestingly, the consistent linear progression in the advection of all four firefronts is noted, even as the firefront advances beyond the saddle point in the given wind field. Furthermore, the observed decrease in slope values with increasing \textit{Da} values across all cases suggests an exponential reduction in group velocity with each increment of \textit{Da}, eventually converging towards an asymptotic value. Additional simulations were performed by extending \textit{Da} to \(10^9\); however, the results are not presented here, as asymptotic convergence was observed, particularly when \(\textit{Da} > 10^3\). This indicates that further increasing \textit{Da} beyond \(10^3\) (minimizing diffusion further) would not yield a significant influence on the deceleration of the firefronts' group velocity. Consequently, an exponential fit regression, expressed as \(a e^{b Da} + c\), is employed to quantify the relationship between the firefront's group velocity and the \textit{Da} number. Here, the coefficients \(a\), \(b\), and \(c\) represent the amplitude, the exponential decay of group velocity, and the asymptotically converged group velocity at infinite \textit{Da} number, respectively. The regression results presented in Table~\ref{tab:group_vel_da_number} show that the four firefronts converge to distinct asymptotic group velocities, reflecting the influence of the applied wind topology. The results emphasize the significant role of the diffusion process in shaping the transient behavior of wildfires, primarily by influencing only the ``magnitude'' of their firefront's group velocity while leaving the ``direction'' of wildfire propagation unaltered.

%Notably, the consistent reduction in the group velocity concerning the increasing \textit{Da} number is characterized by an average exponential decay rate of $-0.943$

\setlength{\tabcolsep}{1.5em}
\begin{table}   
\centering
\begin{tabular}{|c|c|c|c|c|}
\hline
Coefficients & Right front & Left front & Top front & Bottom front \\ \hline
$a$          & $7.575$     & $7.319$    & $7.329$   & $7.527$      \\ \hline
$b$          & $-0.935$    & $-0.958$   & $-0.947$  & $-0.931$     \\ \hline
$c$          & $3.619$     & $3.704$    & $3.318$   & $3.524$      \\ \hline
\end{tabular}
\caption{Characterizing the time-averaged group velocities of the firefronts for the initialized Heaviside firefront under a saddle-type steady wind velocity field over a uniformly distributed fuel bed, using exponential fitting, for different \( Da \in (1, 10, 10^2, 10^3, 10^4,  10^5) \), \( \Phi = 1.0 \), and \( \epsilon = 0.03 \).}
\label{tab:group_vel_da_number}
\end{table}
\subsubsection{Role of Convection--$\Phi$ number}

\begin{figure}[h!]   
    \centering
    \includegraphics[width=1.0\textwidth]{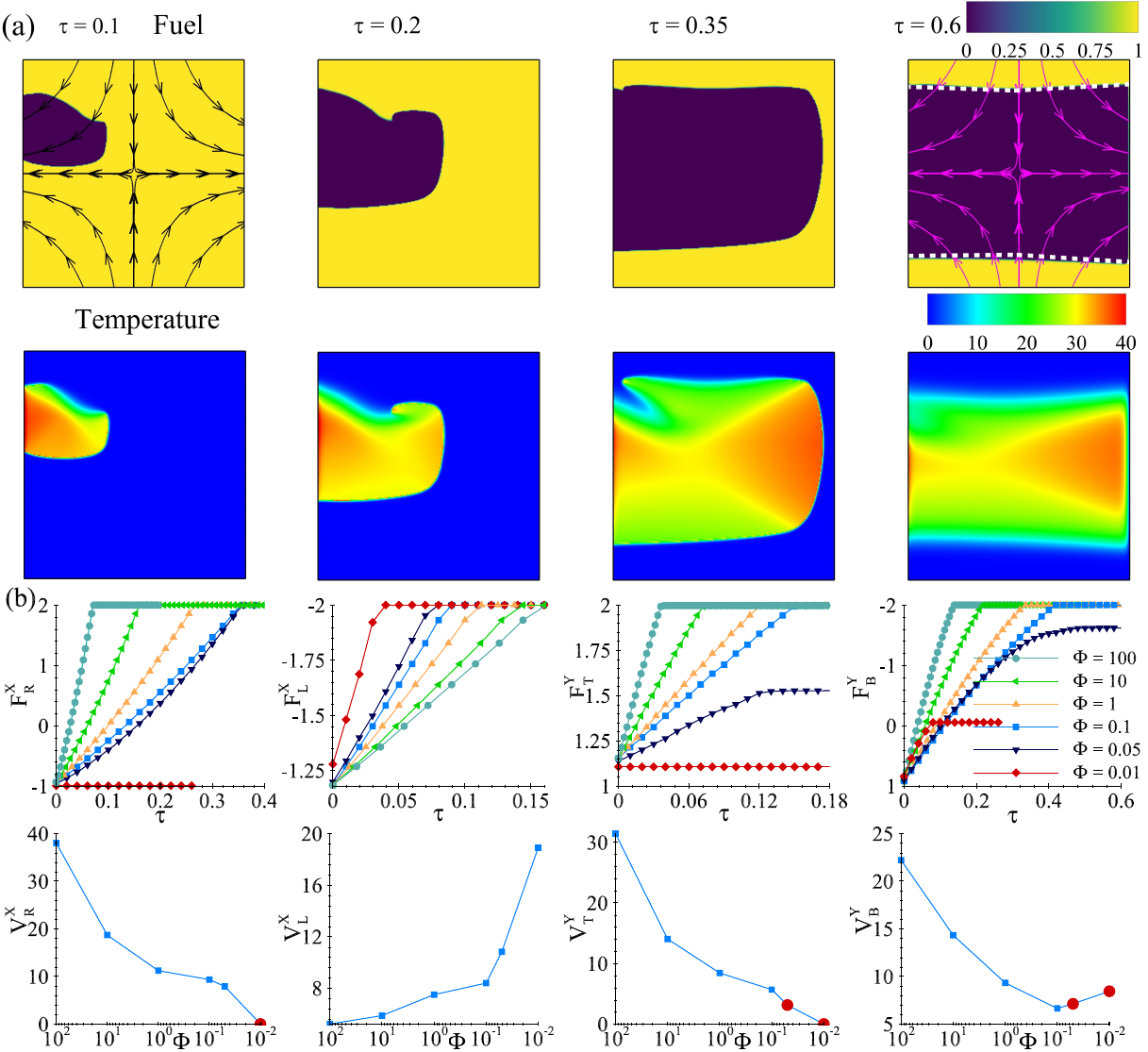}
    \caption{Spatio-temporal evolution of the Heaviside firefront under a saddle-type steady wind velocity field over a uniformly distributed fuel bed, computed across a range of \(\Phi\) values \( \in (100, 10, 1, 0.1, 0.05, 0.01) \), with \(Da = 10^3\) and \(\epsilon = 0.03\). (a) Fuel and temperature fields at various time instants, with the top and bottom rows, respectively, computed for \(\Phi = 0.05\). Streamlines are superimposed at \(\tau = 0.1\) and \(\tau = 0.6\). At \(\tau = 0.60\), the dotted white curve highlights the local neutral curve, indicating the stalled top and bottom firefronts over the fuel bed. (b) Instantaneous spatial advection of the firefronts and their corresponding time-averaged group velocities, depicted in the top and bottom rows, respectively. The red dots highlight the group velocity of stalled firefronts along their respective advecting directions.}
    \label{fig:phi_number}
\end{figure}

To investigate the influence of convection on wildfire propagation,  \(\Phi\)  is varied across values of 100 to 0.01, indicative of increasing wind velocity with decreasing \(\Phi\). \textit{Da} number of \(10^3\) is selected to minimize the diffusion dominance along with \(\epsilon=0.03\). The spatio-temporal evolution of the Heaviside firefront at various time instants is depicted in Fig. \ref{fig:phi_number}a, illustrating both the fuel and temperature fields in the top and bottom rows, respectively, computed for $\Phi = 0.05$. A significant departure from the last section arises here, where the observed fuel bed does not undergo complete consumption. Instead, the advection of the firefronts and the corresponding fuel consumption rate are influenced by imposed stable/unstable manifolds. The attracting streamlines propel both the left and right firefronts forward, effectively encouraging complete fuel consumption along their trajectory. Conversely, the repelling streamlines exert a decelerating effect on the top and bottom firefronts, inducing a phenomenon akin to ``stalling'', where the fire's progression is halted, leaving unburnt fuel remnants. This intriguing observation prompts the identification of the spatial location of the stalled top and bottom firefronts, marked by a white dotted curve at \(\tau = 0.6\) over the fuel bed in Fig.~\ref{fig:phi_number}a. Furthermore, at this spatial location, one observes that the convection, diffusion, and reaction contributions balances with each other (the right-hand side of Eq.~\ref{eq:new_non_dim_eq_scal}), hence both the temperature temporal rate \((\partial \overline{T}/\partial \tau)\) and fuel burn rate \((\partial \beta/\partial \tau)\) reach zero. This notable equilibrium delineates the spatial boundary of the stalled firefronts as a ``local neutral curve'', beyond which the fire's progression ceases and cooling gradually occurs over time as all the fuel has been consumed. 

From the instantaneous positions of the firefronts, shown in Fig. \ref{fig:phi_number}b for different \(\Phi\) values, one can calculate the exact initial stalling time by identifying when the instantaneous group velocity (local slope) reaches zero before the firefront reaches the domain boundary. For the \(\Phi = 0.1\) case, all firefronts advance, consuming all available fuel and reaching the domain boundary, indicating no stalling of firefronts. Conversely, stalling is observed for convection-dominant \(\Phi\) values of 0.05 and 0.01, even for lower values (results are not shown here). Interestingly, in the case of the lowest chosen \(\Phi\) value of 0.01, the initialized firefront burns the fuel only in the second quadrant of the domain, primarily through the advection of the left firefront, while the right, top, and bottom firefronts stall despite advancing beyond the saddle fixed point. Stalling here indicates that the front does not reach the boundary. We computed the time-averaged group velocities of the firefronts with the eventual stalled one indicated by a red dot in the bottom row of Fig. \ref{fig:phi_number}b. A zero-valued time-averaged group velocity signifies the immediate stalling ($\tau \geq 0$) of firefronts, while the cases that stalled with a non-zero time-averaged group velocity had initial advancements followed by eventual stalling. These discernible insights underscore the substantial impact of the convection process on both the magnitude and direction of the firefront's group velocity, while the diffusion process, though primarily decreasing its magnitude, exerts a less pronounced influence than convection and does not alter its direction.

\subsubsection{Role of Reaction -- $\epsilon$ parameter}
\begin{figure}[h!]   
    \centering
    \includegraphics[width=1.0\textwidth]{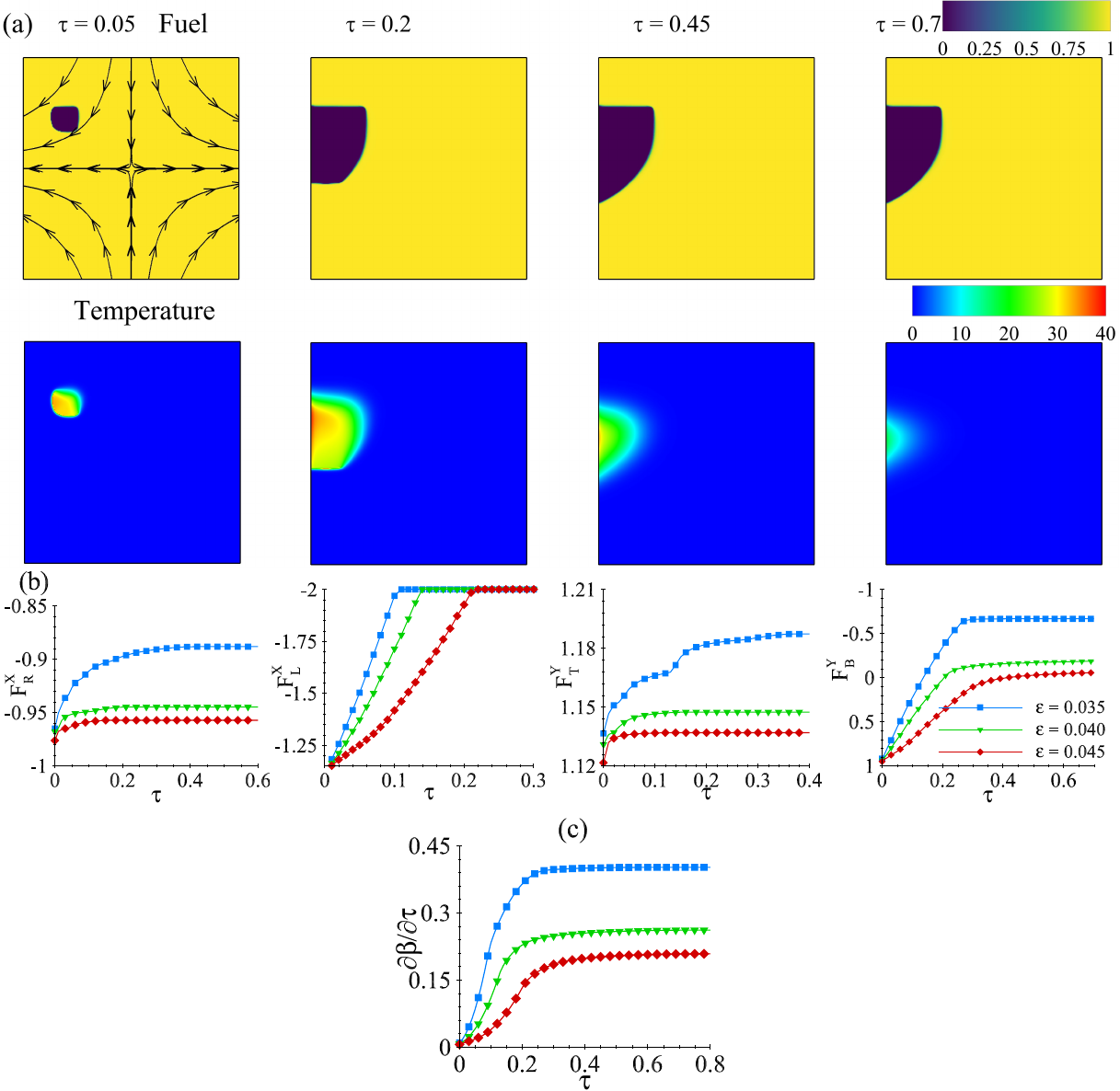}
    \caption{Spatio-temporal evolution of the Heaviside firefront under a saddle-type steady wind velocity field over a uniformly distributed fuel bed, computed across varying \(\epsilon\) values \((0.35, 0.40, 0.45)\), with \(Da = 10^3\) and \(\Phi = 0.05\). (a) Fuel and temperature fields at different time instants, with the top and bottom rows corresponding to each field, respectively, calculated for \(\epsilon = 0.035\). Streamlines are superimposed at \(\tau = 0.05\) over the fuel bed. (b) Instantaneous spatial advection of the four firefronts is plotted. (c) Instantaneous fuel burning rate over time is plotted.}
    \label{fig:eps_number}
\end{figure}
Next, we present how reaction influences transient wildfire behavior by varying $\epsilon$ values (0.35, 0.40, 0.45) while selecting a convection-dominant $\Phi$ value of 0.05 and ensuring minimal diffusion dominance with \textit{Da} set to $10^3$. Fig.~\ref{fig:eps_number} illustrates the spatio-temporal evolution of the Heaviside firefront at various time instances for $\epsilon = 0.035$, with Fig. \ref{fig:eps_number}a showing the fuel and temperature fields in the top and bottom rows, respectively, and Fig. ~\ref{fig:eps_number}b depicting the instantaneous locations of the four advecting firefronts, alongside Fig. ~\ref{fig:eps_number}c, which presents the instantaneous fuel burning rates for the different $\epsilon$ values. Similar to the convection-dominated scenarios, the findings depicted in Fig.~\ref{fig:eps_number}b indicate the stalling of the advecting firefronts in all three $\epsilon$ simulation cases, except for the left firefront, which is driven by the unstable manifold. In particular, from Eq.~\ref{new_non_dim_eq}(a), it is important to recognize the role of $\epsilon$--the reaction parameter--not only in the reaction term but also in the diffusion term, significantly influencing the nonlinearity in the diffusion process. In other words, an increase in $\epsilon$ also promotes nonlinear diffusion, consequently resulting in nonlinear propagation of both left and bottom firefronts, with the instantaneous group velocity exhibiting less linear behavior over time, particularly notable in cases with higher $\epsilon$ values, as illustrated in Fig.~\ref{fig:eps_number}b. Moreover, these increments also induce a decelerating effect on the advection of top and right firefronts, reflecting the simplified fuel burning rate formulation (Eq.~\ref{dim_eq}). Further insight from Fig. \ref{fig:eps_number}c, illustrating the instantaneous computation of \(\partial \beta/\partial \tau\), sheds light on a consistent impediment to the fuel consumption rate due to the decelerated advection of the firefronts, especially as the inverse activation energy (\(\epsilon\)) increases. We remark that this counter-intuitive observation arises from the exponential factor present in the reaction term, which is multiplied by \(\epsilon\) in Eqs. \ref{dim_eq} and \ref{eq:new_non_dim_eq_scal}. 

%However, our current understanding of the influence of reactions remains limited by the simplified fuel model proposed by \citet{asenio}. Incorporating the intricate fuel composition model into the newly derived non-dimensional wildfire equation may unveil deeper insights into the complex dynamics of wildfires, particularly in scenarios dominated by convection, where the reaction factor plays a pivotal role.

\subsection{Transient wind velocity: Double gyre flow}
Now we look into the transient reactive flow dynamics of the physics-based CDR model by investigating their complex interaction with unsteady wind velocity topology, using the double gyre flow, a typical benchmark in environmental flows. Double gyre flow is characterized by a pair of two counter-rotating vortices, as shown in Fig.~\ref{fig:wind_velocity}b. The double gyre flow is constructed by formulating wind velocity components as
\begin{subequations}
\begin{align}
u &= -A_m \sin(\pi f(\xi, \tau)) \cos(\pi \eta) \\
v &= A_m \cos(\pi f(\xi, \tau)) \sin(\pi \eta) \frac{df}{d\xi}
\end{align}
\end{subequations}
% and $\frac{df}{dx} = 2 \lambda \sin(2\pi \widehat{F}\tau) x + \left[1 - 2\lambda \sin(2\pi \widehat{F} \tau)\right]$. 
where \( f(\xi, \tau) = \lambda \sin(2\pi\Omega\tau) \xi^2 + [1 - 2\lambda \sin(2\pi\Omega\tau)] \xi \), with \(\lambda\) the perturbation amplitude representing how far the dividing streamline perturbs from the initial position, \(\Omega\) the oscillation frequency indicating the rate of this perturbation, and \(A_m\) the velocity magnitude. The wind oscillation frequency, normalized by the total fuel consumption timescale \(\overline{t}_r\) (the time required to completely burn all the given fuel), is represented by the Strouhal number \(\textit{St} = \overline{t}_r \Omega\), quantifying the interaction between fuel reaction and flow oscillation timescales. Consistent parameters across simulations include a rectangular domain of \(2 \times 1\) along the \(\xi\) and \(\eta\) axes, a \(256 \times 128\) grid, \(\beta = 1\), a time step of \(\Delta \tau = 10^{-8}\), \(q = 1\), \(\alpha = 10^{-3}\), \(\overline{T}_{\text{pc}} = 3\), \(\overline{\kappa} = 0.1\), and an initial Heaviside square firefront temperature \(\overline{T} = 31\) at \((1.0, 0.5)\) with a side length of \(0.1\) along the dividing streamline. Simulations employ a normalized maximum wind speed \(|\overrightarrow{\mathbf{w}}| = 1\) and LAD coefficient \(C\mu = 0.75\) to examine transient wildfire dynamics under convection-dominant and low-diffusion scenarios, defined by \(\Phi = 10^{-3}\), \(Da = 10^3\), and \(\epsilon = 0.03\). Given the transient wind topology, instantaneous FTLE fields were computed both forward and backward in time over an integration period of \(20\), allowing for the extraction of repelling and attracting Lagrangian coherent structures (rLCS and aLCS) at a given time \(\tau\).

\subsubsection{Role of wind oscillation - St Number}
\begin{figure}[h!]   
    \centering    
    \includegraphics[width=1.0\textwidth]{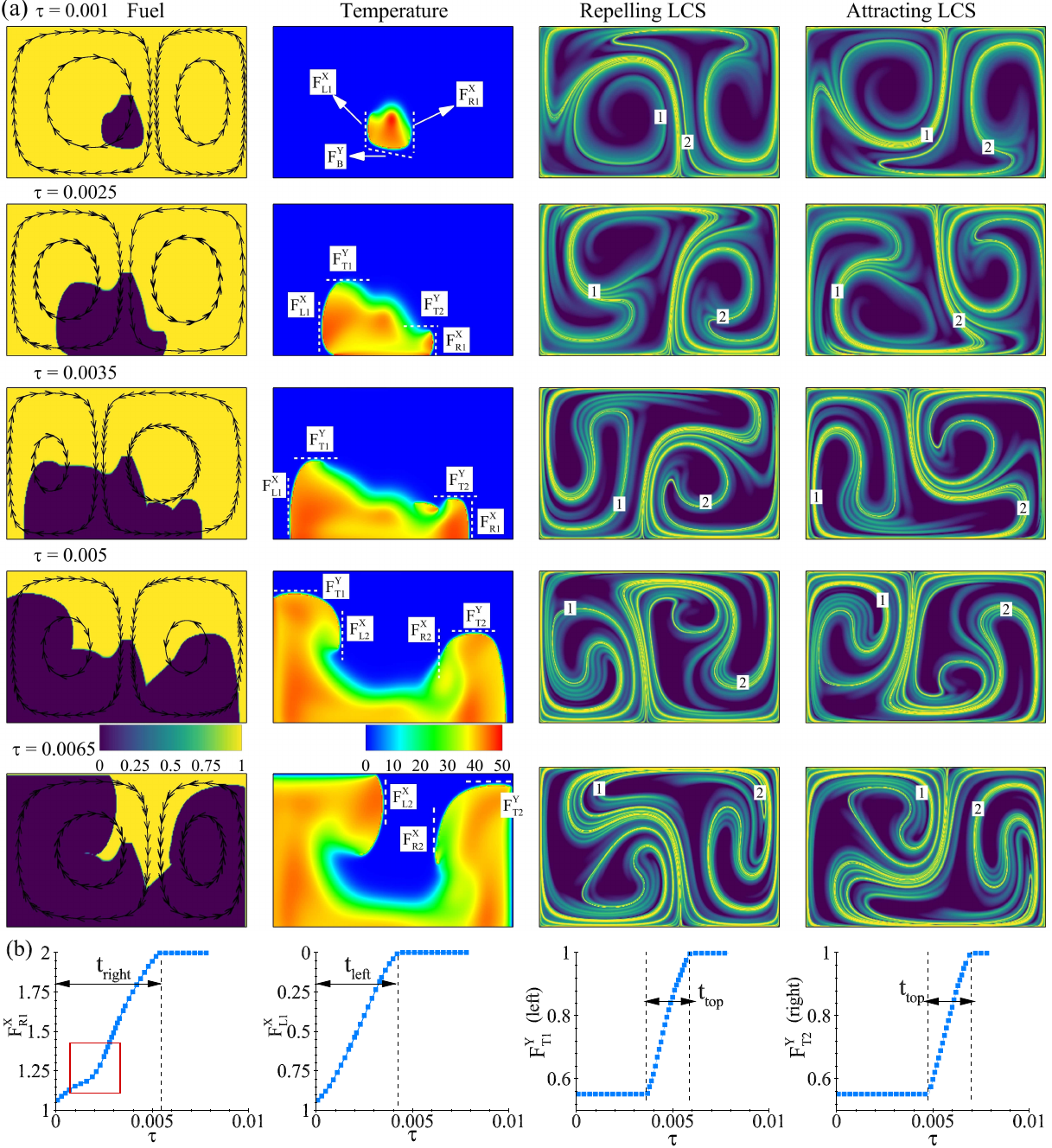}
    \caption{Spatio-temporal evolution of the Heaviside firefront under a double gyre wind velocity field, characterized by parameters \(A_m = 0.1\), \(\textit{St} = 2\), and \(\lambda = 0.25\), over a uniformly distributed fuel bed, computed at \( \Phi = 10^{-3} \), \( \textit{Da} = 10^3 \), and \( \epsilon = 0.03 \). (a) The fuel and temperature fields are shown at various time instants, accompanied by the corresponding repelling and attracting LCS plots, with streamlines superimposed over the fuel bed. (b) The instantaneous spatial advection of the four firefronts is plotted.}
    \label{fig:dg_f_200}
\end{figure}

To effectively investigate the interaction between transient wildfires and time-varying wind topology, we select a wind oscillation timescale \((1/\Omega)\) that aligns with the total fuel reaction timescale \(\overline{t}_r\). A transient wildfire simulation with a steady double-gyre wind velocity \((\Omega = 0)\) determined the fuel consumption timescale \(\overline{t}_r \sim 10^{-2}\). Consequently, we select the wind oscillation frequency to be on the order of \(\overline{t}_r\). In Fig.~\ref{fig:dg_f_200}, the spatio-temporal evolution of the Heaviside firefront at various time instants is presented under an unsteady double-gyre wind flow, oscillating at a frequency of \(\textit{St} = 2\), with an oscillation amplitude of \(\lambda = 0.25\) and a velocity magnitude of \(A_m = 0.1\). Fig.~\ref{fig:dg_f_200}a illustrates the fuel and temperature fields along with the corresponding instantaneous rLCS and aLCS results. At \(\tau = 0.001\), the initialized Heaviside firefront is advected left, right, and downward, resulting in three distinct firefronts--the left firefront \(F_{L1}^{X}\), the right firefront \(F_{R1}^{X}\), and the bottom firefront \(F_{B}^{Y}\)--with dotted lines indicating their locations within the temperature field. Upon examining the fuel bed traces, it becomes evident that the left firefront advances considerably further than the right firefront. This is attributed to the influence of the aLCS-1, which specifically drives the advection of the left firefront while exerting no effect on the right firefront. In contrast, the right firefront experiences moderated advection due to the presence of rLCS-2, which repels and decelerates its progression. Although aLCS-2 is anticipated to contribute to the advancement of the right firefront, its influence is diminished by the preceding rLCS-2, which constrains the progress of the right firefront. This deceleration effect is most pronounced while rLCS-2 remains in the upstream region of the right firefront. However, since wildfire transport is not a pure advection problem and LCS cannot form transport barriers, ultimately, the firefront passes rLCS-2, and aLCS-2 begins to accelerate the advection of the right firefront, as observed at \(\tau = 0.0025\). The deceleration action exerted by rLCS-2 is distinctly evident in the tracked instantaneous location plot of \(F_{R1}^{X}\), illustrated in Fig. \ref{fig:dg_f_200}b, where the red box highlights a notable reduction in the instantaneous firefront's group velocity, followed by a subsequent increment corresponding to the traversal of rLCS-2. This differential behavior emphasizes the contrasting roles of aLCS-2 and rLCS-2 in shaping the firefront dynamics. Moreover, the bottom firefront does not advance uniformly downward but is distorted due to the decelerating influence of rLCS-1, as evidenced by the non-parallel alignment of the identified dotted line in the temperature field.

\begin{figure}[h!]   
    \centering    
    \includegraphics[width=1.0\textwidth]{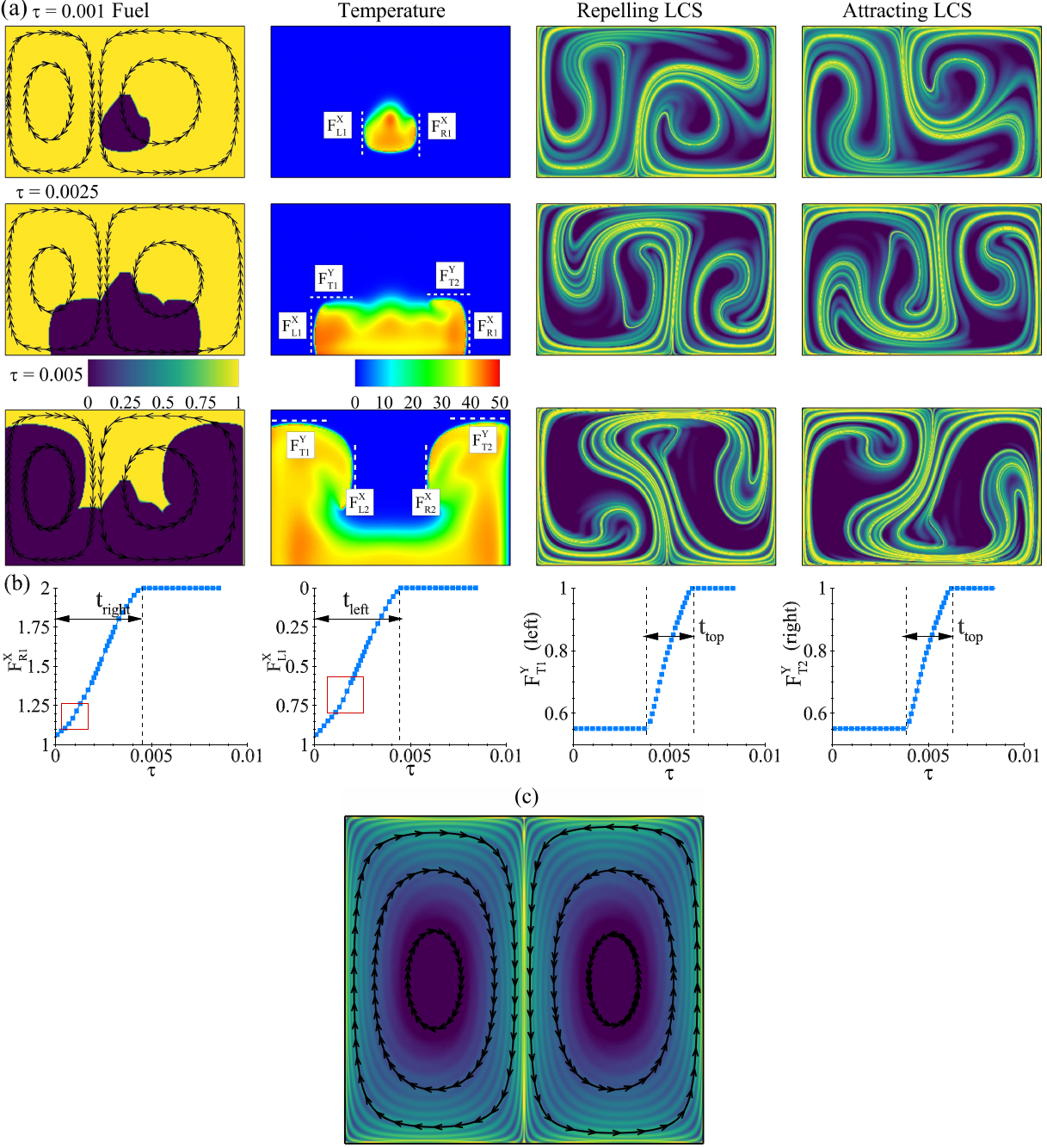}    
    \caption{Spatio-temporal evolution of the Heaviside firefront under a double gyre wind velocity field, characterized by parameters \(A_m = 0.1\), \(\textit{St} = 7.5\), and \(\lambda = 0.25\), over a uniformly distributed fuel bed, computed at \( \Phi = 10^{-3} \), \( \textit{Da} = 10^3 \), and \( \epsilon = 0.03 \). (a) The fuel and temperature fields at various time instants, along with the corresponding repelling and attracting LCS plots, with streamlines superimposed over the fuel bed.  (b) The instantaneous spatial advection of the four firefronts is plotted. (c) The FTLE field corresponding to the time-averaged wind velocity fields is superimposed with time-averaged streamlines.}
    \label{fig:dg_f_750}
\end{figure}

At \(\tau = 0.0025\), two additional distinct top firefronts appear alongside the existing right and left firefronts. These newly identified top firefronts are designated as \(F_{T1}^{Y}\) and \(F_{T2}^{Y}\) within the temperature field, resulting from the advection of the left and right firefronts, respectively. It is observed that the advected distances of these two top firefronts differ, revealing an asymmetry in their advection. This asymmetry is attributable to the repelling influence over \(F_{T2}^{Y}\) firefront by rLCS-2, which precedes the aLCS-2 curve. This phenomenon mirrors the disparity observed between the right and left firefronts at the earlier time \(\tau = 0.001\). In addition, the other three firefronts \(F_{L1}^{X}\) and \(F_{T1}^{Y}\) are attracted by the aLCS-1,  while \(F_{R1}^{X}\) is attracted by aLCS-2 curve, leading to more effective progression without noticeable deceleration. By \(\tau = 0.0035\), the accelerating influence of highlighted aLCS curves on the firefronts \(F_{L1}^{X}\), \(F_{R1}^{X}\), and \(F_{T2}^{Y}\)  becomes more pronounced, effectively overshadowing the decelerative effects of the corresponding rLCS curves, allowing each firefront to advance along its respective trajectory. Interestingly, \(F_{T1}^{Y}\) progresses primarily due to diffusion, as neither aLCS nor rLCS exert any significant influence in this region. At a later time  \(\tau = 0.005\), the continued advection of the two top firefronts gives rise to the emergence of additional left and right firefronts, which are labeled \(F_{L2}^{X}\) and \(F_{R2}^{X}\), respectively. In this scenario, all four identified firefronts experience progressive advection, guided by the attractive influence of the highlighted aLCS curves. It is particularly noteworthy that the highlighted rLCS curves exert no significant influence on the firefronts during this phase. A similar advection scenario is observed at \(\tau = 0.0065\), where the firefronts continue to be effectively attracted by the highlighted aLCS curves. The asymmetry between the left firefront \(F_{L2}^{X}\) and the right firefront \(F_{R2}^{X}\) reflects the earlier asymmetry observed in the earlier case of \(F_{T1}^{Y}\) and \(F_{T2}^{Y}\) firefronts, which, in turn, stems from the differential advection of \(F_{L1}^{X}\) and \(F_{R1}^{X}\). In Fig. \ref{fig:dg_f_200}b, the instantaneous advection of the four firefronts is depicted to quantify its spatio-temporal characteristics in relation to the influence of aLCS and rLCS fields over time \(\tau\). Among the four cases presented, only the right firefront demonstrates a noticeable nonlinear instantaneous group velocity due to the traversal of the rLCS across it, while the other firefronts experience progressive advancement throughout time. The observed asymmetry in the firefronts' advancement is quantitatively reflected in the propagation times shown in the corresponding plots. That is, the left firefront advects faster than the right one (\(t_{left} < t_{right}\)), and a similar trend is observed in the other two top firefronts. Further quantitative analysis of this asymmetry will be discussed in the subsequent section.

%Interestingly, at \( \tau = 0.001 \), it is observed that the right firefront advances more than the left firefront, due to the influence of aLCS

Next, the transient behavior of wildfires is investigated under an unsteady double gyre flow oscillating at a high frequency, characterized by a Strouhal number $\textit{St} = 7.5$, employing the same other parameters, including an oscillation amplitude of $\lambda = 0.25$ and a velocity magnitude of $A_m = 0.1$. This case aims to elucidate the wildfire dynamics when the time scales of fuel consumption and wind flow oscillation are disparate. Fig.~\ref{fig:dg_f_750} illustrates the spatio-temporal evolution of the Heaviside firefront under the considered unsteady double gyre flow scenario As anticipated, the advection of all firefronts nearly exhibits symmetric advancement across the fuel bed, since the flow oscillation time is considerably smaller than the fuel consumption time scale, implying that the firefronts do not have sufficient time to respond to the wind oscillations. Consequently, the subsequent emergence of top firefronts (resulting from the advection of right/left firefronts) also displays a nearly symmetric advection. The instantaneous locations of the tracked firefronts, as depicted in Fig. \ref{fig:dg_f_750}a, further substantiate this nearly symmetric advection (\(t_{left} \approx t_{right}\)) over the evolution time $\tau$. Although changes in local group velocity for both the right and left firefronts are evident, as highlighted within the red box panel, the local deceleration/acceleration is minimal compared to the last transient case. The instantaneous FTLE fields and corresponding aLCS and rLCS reveal an asymmetric pattern, which appears to be unrelated to the high-Strouhal number oscillation scenario. Instead, FTLE calculation on the time-averaged wind velocity field, as shown in Fig. \ref{fig:dg_f_750}c is sufficient to characterize the advection of firefronts in this case. Notably, the LCS manifests as a ridge-like structure along the dividing streamline. In the present case, all firefronts adhere to the time-averaged local streamlines, much like the behavior observed in a saddle-type steady wind velocity scenario.

%From the investigation of transient wildfire behavior under the influence of unsteady double gyre flow, a critical question arises: Under what conditions should one utilize instantaneous FTLE rather than time-averaged FTLE for predicting firefront propagation? Furthermore, does the validity of FTLE persist when the wind topology is unsteady? The response is rooted in the robustness of LCS, which remains \textit{valid} irrespective of the transient nature of the wind topology. Specifically, the necessity for instantaneous FTLE computation in understanding wildfire propagation hinges on the degree of asymmetry observed in the advection of firefronts. 

By comparing two Strouhal number cases, it becomes evident that asymmetric advection of firefronts is more pronounced when \(\textit{St} = 2\), where wind oscillation closely aligns with the fuel consumption time scale. In contrast, minimal asymmetry is observed when \(\textit{St} = 7.5\), where the respective time scales are misaligned, indicating that the degree of asymmetry is most notable when \(\textit{St} \rightarrow 1\).  To address this quantitatively, one can project the advection characteristics of the firefronts onto a Bode plot--similar to the approach used by \citet{ducruix2000theoretical} for transient premixed flames--thereby providing an understanding of the wildfire's response to unsteady wind conditions. 

\subsubsection{Dynamic response of wildfires to unsteady wind oscillation}
The dynamic response of wildfires to the imposed unsteady double-gyre wind topology is systematically quantified by projecting the firefronts advection characteristics onto a Bode plot through a transfer function approach. In the Bode plot, a transfer function (TF) is generally defined as $TF=\overline{R}\angle\phi$, where magnitude $(\overline{R})$  denotes the amplitude gain, and the phase angle $(\phi)$ indicates the phase difference of an output and an input response. 

Transfer function of the right firefront $(TF_R)$ is given by
\begin{equation}
    TF_R = \overline{R}_R \angle\phi_R \hspace{2mm};\hspace{2mm}
    \begin{cases}
     \overline{R}_R = \dfrac{V_R^{X}}{V_L^{X}} \\[10pt]
      \phi_R = \left[ \dfrac{t_{right}} {t_{right}\big\rvert_{St = 0}}  - 1 \right] \pi \;.
    \end{cases} 
    \label{tf1}
\end{equation}

Similarly, the transfer function of the left firefront $(TF_L)$ is given by
\begin{equation}
     TF_L = \overline{R}_L \angle\phi_L \hspace{2mm};\hspace{2mm}
    \begin{cases}
      \overline{R}_L = \dfrac{V_L^{X}}{V_R^{X}}  = \dfrac{1}{\overline{R}_R}\\[10pt]
      \phi_L =  \left[ \dfrac{t_{left}} {t_{left}\big\rvert_{St = 0}}  - 1 \right] \pi \;.
    \end{cases}
    \label{tf2}
\end{equation}

\begin{figure}[h!]   
    \centering
    \includegraphics[width=\textwidth,height=\textheight,keepaspectratio]{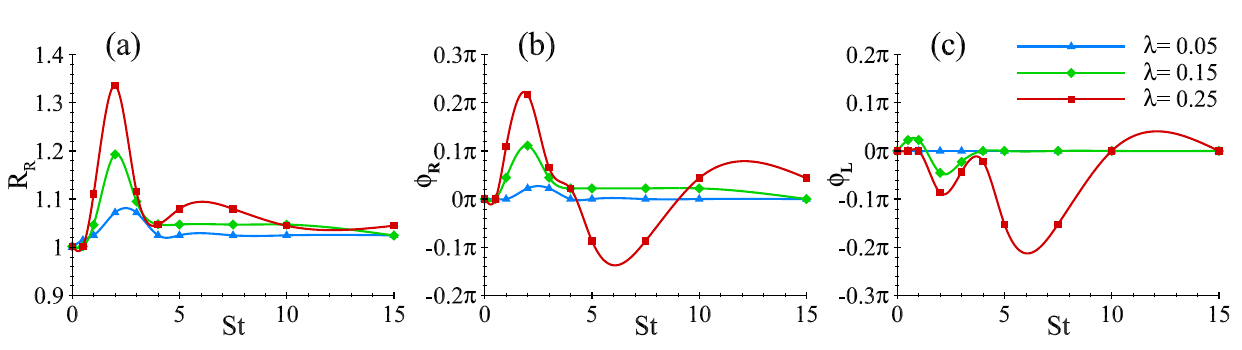}
    \caption{Bode plots -- \((a)\) The magnitude of the $TF_R$, \((b)\) the phase angle of the $TF_R$, and \((c)\)  the phase angle of the $TF_L$ are shown. These results are computed with \(A_m = 0.1\), \(\Phi = 10^{-3}\), \(Da = 10^3\), and \(\epsilon = 0.03\) for different wind oscillation amplitudes $(\lambda)$ and Strouhal numbers.}
    \label{fig:bode_plot}
\end{figure}

It is important to note that the magnitude of the TF quantifies the degree of asymmetry between the advection of the right and left firefronts by calculating their time-averaged group velocity, thereby reflecting the uneven spread of the fire under the influence of the unsteady double-gyre wind. A TF magnitude different from unity indicates asymmetric transport, suggesting that similar asymmetry should be observed in the advection of subsequent formations of other top firefronts, as discussed earlier. Additionally, the phase angle of the TF shows the phase response of the firefronts relative to a baseline scenario characterized by a steady wind topology (\(\textit{St} = 0\)), where a negative phase angle indicates a phase lag and a positive phase angle denotes a phase advance in the response of the firefront progression to wind oscillation. Fig.~\ref{fig:bode_plot} presents the Bode plot illustrating the transient wildfire characteristics of firefronts subjected to an unsteady double gyre flow, demonstrating the transfer function of both right and left firefronts across varying Strouhal numbers and oscillation amplitudes, $\lambda \in (0.05, 0.15, 0.25)$. The $\overline{R}_L$ plot is omitted as it is the inverse of $\overline{R}_R$. Notably, the phase response of both firefronts is not monotonous, reaching zero for certain Strouhal numbers, which indicates the existence of a \textit{phase inversion frequency} in the transient wildfire propagation scenario under unsteady wind conditions, implying a phase-switching phenomenon where the fire response shifts from advance to lag or vice versa, as similarly observed in~\citet{motta2015influence, vik_flow} for unsteady flow control applications.

From Fig.~\ref{fig:bode_plot}a, it is evident that for any given oscillation amplitude $\lambda$, the $\overline{R}_R$ value of the right firefront exhibits a significant deviation from unity and a resonance condition when $1 \lessapprox \textit{St} \lessapprox 4$. Conversely, for small Strouhal numbers and oscillation amplitudes $\lambda$, $\overline{R}_R$ approaches unity, signifying that the effect of wind unsteadiness diminishes. In contrast, for very large Strouhal numbers, the firefront has insufficient time to effectively respond to the oscillations, resulting in $\overline{R}_R$ remaining close to unity with minimal deviation due to the rapid oscillations. Examination of the real part of the transfer function ($TF_R$) in the Bode plot reveals that, under resonance conditions with moderate Strouhal numbers, instantaneous FTLE calculations are essential for precise prediction of wildfire dynamics. Conversely, when the Strouhal number is significantly greater than one ($\textit{St} \gg 1$) and oscillation amplitudes are small ($\lambda \ll 1$), FTLE values based on time-averaged wind velocity are adequate for forecasting wildfire behavior across the fuel bed.

\section{Conclusion}
In the present research, the reactive flow-wildfire transport model proposed by \citet{asenio} has been employed to study the role of flow topology under two benchmark wind flow typologies: steady wind with saddle-type fixed points and unsteady wind characterized by double gyre flow. The key outcomes of this study are enumerated as follows:

\par $(a)$ The wildfire combustion model has been revisited and non-dimensionalized by introducing three distinct time scales that define the convection-diffusion-reaction process. This revision led to the identification of two non-dimensional numbers: the Damköhler number (\textit{Da}) and a newly defined non-dimensional number $(\Phi)$, representing the ratio of the Damköhler number to the Peclet number. This approach contrasts with the conventional non-dimensionalization, which typically incorporates a single time scale, thereby culminating in a revised new non-dimensional wildfire combustion model with additional physical insight.

\par $(b)$ Through scaling analysis, critical determinants of transient wildfire behavior were identified, including the state-neutral curve where the temporal rate of fire temperature is zero for specific combinations of the two non-dimensional numbers. This finding offers a valuable tool for predicting overall wildfire dynamics from initial conditions, thus mitigating the reliance on extensive and computationally intensive simulations.

\par $(c)$ A robust wildfire transport solver has been developed, leveraging CUDA support within a finite difference method framework. The solver employs the upwind compact schemes and implicit-explicit Runge-Kutta methods to resolve both spatial and temporal solutions while incorporating the LAD model to selectively attenuate high-frequency instabilities without compromising accuracy in smooth regions.

\par $(d)$ In the steady saddle-type fixed point wind topology, it was demonstrated that the unstable manifold (attracting LCS) significantly influences firefronts, guiding them along their trajectory. Conversely, the stable manifold (repelling LCS) exerts a decelerating influence, causing the firefronts to stall and inhibiting further propagation, particularly under conditions characterized by a large \textit{Da} and a small $\Phi$, where convection predominates, and diffusion is minimal. However, due to the simplified fuel reaction composition in the \citet{asenio} model, definitive conclusions regarding the reaction component's interaction with both stable and unstable manifolds remain to be investigated in detail.

\par $(e)$ In the context of unsteady double gyre flow, the investigation illustrated that instantaneous FTLE fields--comprising unstable and stable manifolds--are pivotal in governing the advection of firefronts at moderate Strouhal numbers, particularly when resonance occurs with wildfire--wind oscillations. In contrast, at very high Strouhal numbers, where off-resonance conditions prevail, wildfire propagation is more strongly influenced by time-averaged FTLE fields. By projecting the advection characteristics of firefronts onto a Bode plot, it becomes possible to determine the conditions under which either the instantaneous or time-averaged FTLE should be computed for more accurate predictions of wildfire propagation. Furthermore, the existence of \textit{phase inversion frequency} (Strouhal number) has been identified for wildfire propagation when subjected to transient wind conditions.

Although extensive numerical wildfire simulations were conducted, it is important to recognize that the findings of this study are derived from a simplified wildfire model that focuses on reactive flow dynamics, which was specifically chosen to investigate the fundamental relevance of LCS to wind-driven wildfire propagation by resolving only the flame scale. \edit{Additionally, the wildfire model assumes that the radiative flux only acts \textit{locally} by treating the medium as optically thick, adopting the Rosseland approximation,  wherein the radiative heat flux is expressed as a nonlinear diffusion. Alternatively, this assumption can be relaxed by employing a global radiation model, as proposed in~\citet{margerit2002modelling, ferragut2007numerical}, which accounts for long-range radiative interactions and, in a more recent method,~\cite{navas2024modeling} incorporates coupled global radiative–wind effects, albeit with increased computational complexity.}Nevertheless, the correspondence of stable and unstable manifolds of wind topology with firefront progression is analyzed within this simplified, constrained framework, it may still serve as a viable basis for extending its applicability to real-world wildfire scenarios. Hence, future research should adopt a more comprehensive approach by employing fully coupled CFD-wildfire methodologies to further advance the understanding of wildfire dynamics and the correspondence of LCS in more complex and realistic settings. In addition, incorporating detailed vegetation distributions, including firebreaks, and accounting for actual terrain topography—such as both uphill and downhill scenarios—is essential for accurately capturing terrain effects on wildfire behavior. Ultimately, these additional effects will not only enhance predictive models but also provide deeper insights into the intricate mechanisms governing wind-driven wildfire propagation.

\section*{Acknowledgment}
We gratefully acknowledge the NSF EAGER grant (Award No.~2330212) from the Combustion and Fire Systems program. Additionally, co-author AT would like to acknowledge partial support from the Industry Advisory Board of the NSF-IUCRC Wildfire Interdisciplinary Research Center (WIRC) at San Jos\'e State University (Award No.~2113931).

\section*{Declaration of Interests}
The authors report no conflict of interest.

\section*{Data Availability}
\edit{
The developed GPU-accelerated Wildfire Transport FDM Python Solver is publicly available on GitHub: \href{https://github.com/siva-viknesh/Wildland-Fire-Dynamics}{\texttt{https://github.com/siva-viknesh/Wildland-Fire-Dynamics}}.

}
\appendix

\edit{
\section{FDM Solver Algorithm}
\label{appendix:solver_alog}
This section presents the FDM solver algorithm developed to simulate wildfire propagation, incorporating the selected physics-based CDR combustion model,~\cite{asenio}. The numerical solver employs the optimized fifth-order upwind compact scheme (OUCS2) for the convection term and a second-order central difference scheme (CD2) for both physical diffusion and localized artificial diffusion components. Temporal integration is performed using the IMEX-RK method, which enables simultaneous integration of stiff ($\overline{\boldsymbol{T}}^{(i)}$) and non-stiff ($\widehat{\boldsymbol{T}}^{(i)}$) temperature and fuel fields. Two wind topology scenarios are considered: (i) a steady wind field characterized by a saddle-type fixed-point structure, and (ii) a transient wind field governed by a double-gyre configuration, introducing time-varying wind dynamics parameterized by the Strouhal number. The complete FDM solver algorithm is detailed in Algorithm~\ref{alg:SteadyWind}.

For the steady wind case, the algorithm begins by initializing the temperature and fuel fields, denoted as \( \boldsymbol{T}^0(x) \) and \( \boldsymbol{\beta}^0(x) \), respectively. Additionally, physical and numerical parameters, relevant non-dimensional numbers, and the wind velocity field are prescribed. At each time step \( n \), the solver first enforces the boundary condition representing zero-flux constraints at the domain boundaries. Subsequently, for each IMEX-RK stage \( i \), the temperature field is partitioned into an explicit (non-stiff) component \( \widehat{\boldsymbol{T}}^{(i)} \) and an implicit (stiff) component \( \overline{\boldsymbol{T}}^{(i)} \), while the fuel field \( \boldsymbol{\beta}^{(i)} \) is updated implicitly. At each temporal stage \( i \), the slope \( F_i^\beta \) associated with the fuel field is computed via a closed-form expression obtained from the discretized stiff fuel equation. The temperature slope \( F_i^T \) is then evaluated by summing the following four contributions:

\begin{itemize}
    \item \textbf{Convection} (\( \mathcal{C}_i^n \)): Evaluated using the OUCS2 operator, this term captures the advective transport of firefronts driven by the prescribed wind field and is treated explicitly.

    \item \textbf{Diffusion} (\( \mathcal{D}_i^n \)): Incorporates both stiff and non-stiff temperature components. The thermal diffusivity \( K(\widehat{\boldsymbol{T}}^{(i)}) \) is computed from the non-stiff temperature component, while spatial derivatives of the stiff temperature are approximated using the CD2 scheme. Here, the first derivative form of the CD2 scheme is employed twice, one after the other, due to its conservative form representation.

    \item \textbf{Reaction and Natural Convection} (\( \mathcal{R}_i^n \)): Represents the nonlinear source and sink terms. The reaction term arising from fuel consumption is treated implicitly, while the natural convection component is handled explicitly.

    \item \textbf{Localized Artificial Diffusion} (\( \mathcal{L}_i^n \)): The spatially varying artificial diffusion coefficient \( \mu^n \) is determined using a biharmonic operator acting on the non-stiff temperature, followed by a Heaviside function that isolates regions near the firefront. This coefficient is scaled by a tunable parameter \( C_\mu \). Then the localized artificial term is computed by applying a Laplacian operator to the non-stiff temperature, scaled by \( \mu^n \).
\end{itemize}
The total temperature slope \( F_i^T \) is then constructed by accounting for all four contributions and is solved using the iterative Newton–Raphson method with a convergence tolerance of \(10^{-12}\). Here, \( a_{ij} \) and \( \tilde{a}_{ij} \) represent the coefficients in a standard Butcher tableau, used to calculate internal temporal stages of the IMEX-RK scheme setting. Upon completing all temporal stages, the temperature and fuel fields are updated through a weighted combination of the computed stage slopes, thereby advancing the solution in time. For the transient wind topology case, the only modification to the procedure involves the computation and update of the instantaneous wind velocity field characterized by the Strouhal number, at each temporal marching.
}

\begin{algorithm}
\edit{
\caption{\textbf{Wildfire CDR Algorithm}}
\label{alg:SteadyWind}
\begin{algorithmic}
    \State \textbf{Input:}
    \Statex \quad Initial fields: $\boldsymbol{T}^0(x)$, $\boldsymbol{\beta}^0(x)$
    \Statex \quad Time step: $\Delta \tau$, number of steps: $N_\tau$
    \Statex \quad Parameters: $\overline{\kappa}$, $\alpha$, $\epsilon$, $q$, $\overline{T}_{pc}$, $C_\mu$
    \Statex \quad Non-dimensional numbers: $\Phi$, $Da$
    \Statex \quad Temporal stages: s
    \Statex \quad Velocity field: $\overrightarrow{\mathbf{w}}(\xi, \eta)$

    \State \textbf{Operator Construction:}
    \Statex \quad First derivative operators: $\nabla$ (OUCS2), $\overline{\nabla}$ (CD2)
    \Statex \quad Laplacian: $\nabla^2$ (CD2), Biharmonic: $\nabla^4$ (CD2)

    \State \textbf{Temporal Marching:}
    \For{$n = 0$ to $N_\tau - 1$}
        \State \textbf{Apply Boundary Condition} : $
        \left[\overrightarrow{\mathbf{w}}\overline{T}^n - K(\overline{T}^n) \nabla \overline{T}^n\right] \cdot \hat{n} = 0$
        \State \textbf{Compute stage values for} $2 \leq i \leq s$:
        \Statex \qquad $\widehat{\boldsymbol{T}}^{(i)} \gets \boldsymbol{T}^n + \Delta \tau \displaystyle\sum_{j=1}^{i-1} \tilde{a}_{ij} F_j^T$,  \quad $\overline{\boldsymbol{T}}^{(i)} \gets \boldsymbol{T}^n + \Delta \tau \displaystyle\sum_{j=1}^{i-1} a_{ij} F_j^T$, \quad $\boldsymbol{\beta}^{(i)} \gets \boldsymbol{\beta}^n + \Delta \tau \displaystyle\sum_{j=1}^{i-1} a_{ij} F_j^\beta$
        
        \State \textbf{Solve for  $F_i^\beta$:}
        \begin{equation*}
        F_i^\beta \gets -\frac{\epsilon}{q} \frac{\boldsymbol{\beta}^{(i)} s(\widehat{\boldsymbol{T}}^{(i)})^{+} \exp(\widehat{\boldsymbol{T}}^{(i)}/[1 + \epsilon \widehat{\boldsymbol{T}}^{(i)}])}{\left[1 + (\epsilon/q) \Delta \tau a_{ii} s(\widehat{\boldsymbol{T}}^{(i)})^{+} \exp(\widehat{\boldsymbol{T}}^{(i)}/[1 + \epsilon \widehat{\boldsymbol{T}}^{(i)}])\right]}
        \end{equation*}

        \State \textbf{Solve a linear system for each $F_i^T$:}
        \Statex \qquad \qquad \textbf{Convection:}
        \begin{equation*}
        \mathcal{C}_i^n \gets \frac{\overrightarrow{\mathbf{w}}}{\Phi} \cdot \nabla \widehat{\boldsymbol{T}}^{(i)}
        \end{equation*}

        \Statex \qquad \qquad \textbf{Diffusion:}
        \begin{equation*}
        \mathcal{D}_i^n \gets \frac{1}{Da} \, \overline{\nabla} \cdot \left[ K(\widehat{\boldsymbol{T}}^{(i)}) \, \overline{\nabla} \left( \overline{\boldsymbol{T}}^{(i)} + \Delta \tau a_{ii} F_i^T \right) \right]
        \end{equation*}

        \Statex \qquad \qquad\textbf{Reaction and Natural Convection:}
        \begin{equation*}
        \qquad \mathcal{R}_i^n \gets s(\widehat{\boldsymbol{T}}^{(i)})^+ \left[ \boldsymbol{\beta}^{(i)} + \Delta \tau a_{ii} F_i^\beta \right] \exp(\widehat{\boldsymbol{T}}^{(i)}/[1 + \epsilon \widehat{\boldsymbol{T}}^{(i)}]) - \alpha \left[ \overline{\boldsymbol{T}}^{(i)} + \Delta \tau a_{ii} F_i^T \right]
        \end{equation*}

        \Statex \qquad \qquad \textbf{Localized Artificial Diffusion:}
        \begin{equation*}
        \mathcal{L}_i^n \gets \mu^n \nabla^2 \widehat{\boldsymbol{T}}^{(i)}; \quad \mu^n \gets C_\mu \, \mathbf{H_F}\left[ | \nabla^4 \widehat{\boldsymbol{T}}^{(i)} | \right]
        \end{equation*}
        \Statex \qquad \qquad \textbf{Total slope at stage  $i$:}
        \begin{equation*}  \qquad F_i^T \gets \mathcal{D}_i^n + \mathcal{R}_i^n - \mathcal{C}_i^n + \mathcal{L}_i^n
        \end{equation*}
    \EndFor
    \Statex \qquad $\boldsymbol{T}^{n+1} \gets \boldsymbol{T}^n + \Delta \tau \displaystyle\sum_{j=1}^s b_j F_j^T$,  \quad $\boldsymbol{\beta}^{n+1} \gets \boldsymbol{\beta}^n + \Delta \tau \displaystyle\sum_{j=1}^s b_j F_j^\beta$
    \Statex \quad Marched solution: $\boldsymbol{T}^{n+1}$, $\boldsymbol{\beta}^{n+1}$ at $\tau^{n+1} = \tau^n + \Delta \tau$
\end{algorithmic}
}
\end{algorithm}

\section{Numerical FDM solver validation}
\label{appendix:validation}
\begin{figure}[h!] 
    \centering
    \includegraphics[width=0.8\textwidth]{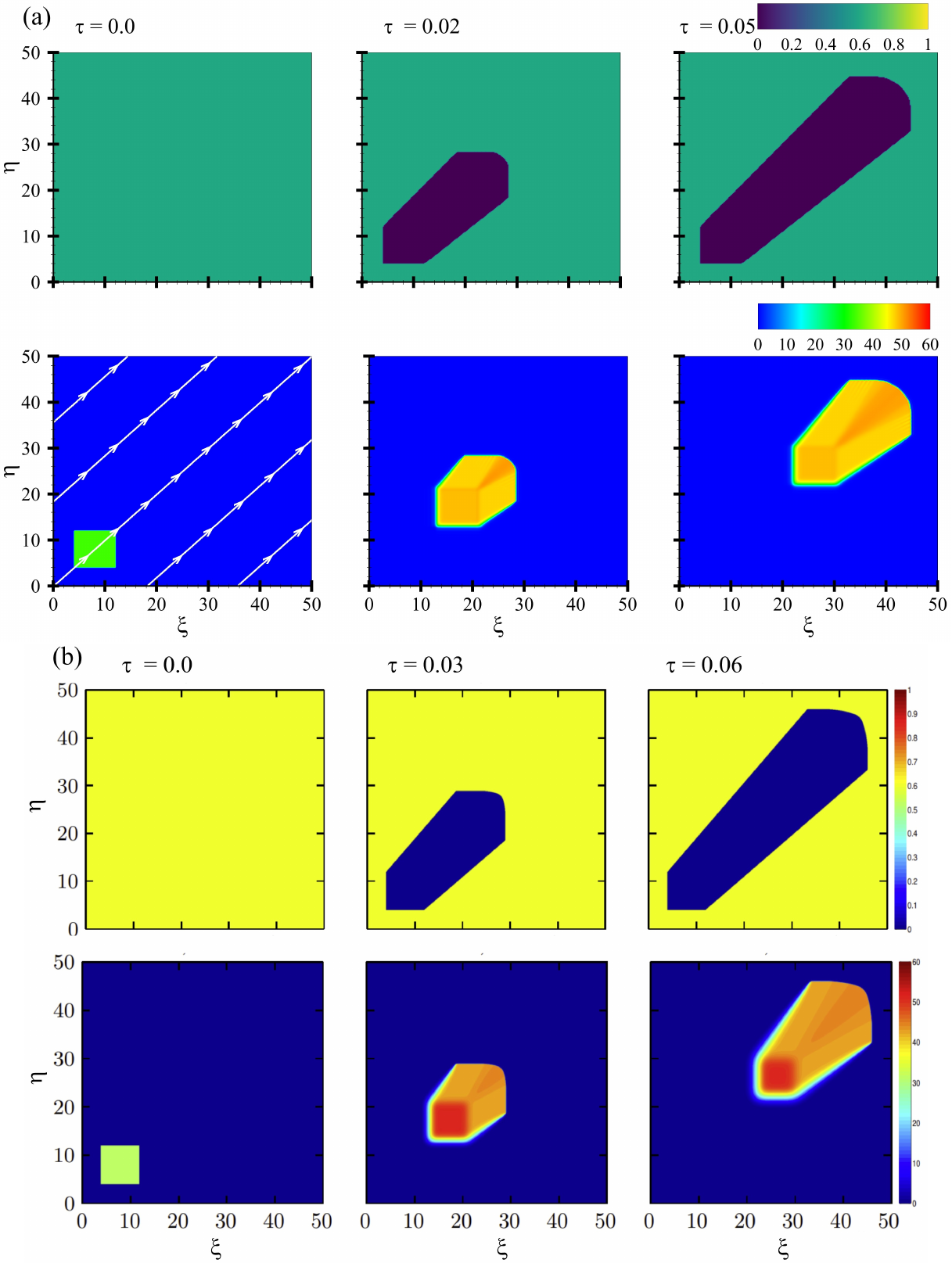}
    \caption{Comparison of the spatio-temporal evolution of the Heaviside firefront over uniformly distributed fuel, (a) computed using our FDM solver and compared against (b) results from~\citep{imex}. Both fuel and temperature distributions are depicted at various time instants in the top and bottom rows, respectively. Panel (b) in the figure adapted from~\citet{imex} with permission. Copyright MDPI 2020.
    }
    \label{fig:valid}
\end{figure}

In this section, we validate our solver using a benchmark 2D perfect initial Heaviside firefront over a uniform fuel distribution scenario, as presented in \citet{imex}. Resolving the advecting Heaviside firefront over a long integration time poses a significant numerical challenge, as it excites all possible frequencies, including spurious ones associated with Gibbs' oscillation in a Fourier spectrum.  Despite this challenge, the OUCS2 compact scheme adeptly captures the spatiotemporal evolution of the Heaviside firefront effectively, even \textit{without} incorporating the LAD model for the considered wildfire scenario (Fig.~\ref{fig:valid}). We maintain consistency by adhering to the parameters in~\citet{imex}: a square domain size of \( 50 \), \( \epsilon = 3 \times 10^{-2} \), \( q = 1 \), \( \alpha = 10^{-3} \), \( \overline{T}_{\text{pc}} = 3 \), \(\overline{\kappa} = 0.1\), initial firefront temperature \( \overline{T} = 31 \), wind speed \( |\overrightarrow{\mathbf{w}}| = 425 \) with a direction of \( 45^{\circ} \), $C_\mu = 0$, and \( Da = \Phi = 1 \). Furthermore, with the initial firefront precisely positioned at coordinates \( (8.0, 8.0) \) and a square side length of 4, and the fuel uniformly initialized at 0.6 across the spatial domain, the simulation is conducted with a time step of \( \Delta \tau = 10^{-7} \) over a grid size of \(256 \times 256\). The results depicted in Fig. \ref{fig:valid}, illustrate the spatio-temporal evolution of the Heaviside firefront at different time instants, demonstrating good agreement with the \citet{imex} results. Discrepancies may arise from using a more accurate IMEX-RK3 time integration and  OUCS2 schemes in our work, contrasting with the employment of IMEX-RK2 scheme and highly dissipative WENO scheme in~\citet{imex}.   

%With the solver successfully validated, we now proceed to investigate the behavior of the firefront in wildfire scenarios under various complex steady (involving a saddle-type fixed point) and transient (double gyre flow) wind velocity fields, as discussed in the subsequent sections.

\edit{
\section{Methodological Comparison of Wildfire Models}
\label{appendix:method_comp}
In this section, a comparative analysis is presented between three wildfire modeling approaches: the semi-empirical model introduced by~\cite{rothermel1972mathematical}, the CDR wildfire model utilized in this work, and the fully coupled wildfire CFD method, FIRETEC. The comparison is conducted by examining the rate of spread ($R_s, m/s$) predictions for a benchmark wildfire scenario involving fireline propagation over a flat terrain under unidirectional wind conditions. Specifically, the performance of the CDR wildfire model is evaluated against the study of~\cite{pimont2012coupled}, which provides a comparison between FIRETEC simulations and the Rothermel model across various slope and wind conditions. Mathematically, the semi-empirical Rothermel model expresses the rate of spread as
\begin{equation}
R_s = R_0 (1 + \phi_u + \phi_s) \;,
\label{Rothermel_eq}
\end{equation}
where $R_0$ denotes the rate of spread under no-wind and zero-slope conditions, $\phi_u$ accounts for the influence of wind (upwind or downwind), and $\phi_s$ represents the contribution due to terrain slope. In~\cite{pimont2012coupled}, the simulations were conducted within a cuboidal computational domain with a square base of $320 \times 320$~m, a vertical extent of 615~m, comprising 60 simulations, spanning three wind velocities (1, 5, 12~m/s), two initial fireline lengths (20~m and 50~m), and 10 different slope conditions. Their results indicated that the dependence of spread rate on different wind and slope conditions is neither purely multiplicative nor strictly additive, deviating from the semi-empirical formulation given by Eq.~\ref{Rothermel_eq}. Furthermore, a nonlinear coupling between wind, slope, and fireline length was identified as a key factor influencing the rate of spread. For the Rothermel model calculation,  the coefficient value of $R_0$ and the wind-related coefficient $\phi_u$ were computed using the FIRETEC simulation data. In contrast, the slope-related coefficient $\phi_s$ was evaluated using the expression given in~\cite{rothermel1972mathematical}.

For the present comparison, we consider \textit{Quercus coccifera}, one of the fuels analyzed in~\cite{pimont2012coupled}, whose properties are summarized in Table~\ref{tab:fuel_prop}. The rate of spread ($R_s$) results are extracted from Fig.~2a of~\cite{pimont2012coupled}, which presents the comparative predictions across varying wind and slope conditions; however, to remain consistent with the primary focus of this comparative analysis, only the zero-slope cases (\( \phi_s = 0 \)) and three wind velocity scenarios for initial fireline length of 50 m are selected. Our CDR wildfire simulations are configured to closely align with the three-dimensional FIRETEC setup, with model parameters and non-dimensional groups computed using the mean values of physical quantities, as listed in Table~\ref{tab:fuel_prop}. It should be noted that such averaging introduces an independent modeling problem, which may affect predictive accuracy.

\begin{table}[h!]
\centering
\small
\renewcommand{\arraystretch}{1.2}
\begin{tabular}{|c|c|c|c|c|c|}
\hline
\textbf{Material} &
  \begin{tabular}[c]{@{}c@{}}$\rho$\\ (kg/m$^3$)\end{tabular} &
  \begin{tabular}[c]{@{}c@{}}$C$\\ (kJ/kg·K)\end{tabular} &
  \begin{tabular}[c]{@{}c@{}}$k$\\ (W/m·K)\end{tabular} &
  \begin{tabular}[c]{@{}c@{}}$A$\\ (s$^{-1}$)\end{tabular} &
  $\epsilon$ \\ \hline
Quercus coccifera & 700   & 1.3–1.5   & 0.12–0.20   & $10^7$–$10^{12}$ & 0.016–0.038 \\ \hline
Air               & 1.293 & 1.005     & 0.025–0.030 & –                & –           \\ \hline
Mean values       & 100   & 1         & 0.1         & $10^{10}$           & 0.03        \\ \hline
\end{tabular}
\caption{Material properties employed in the FIRETEC wildfire simulations, as reported in~\cite{pimont2012coupled} and the references therein, along with the representative mean values adopted in the present comparative analysis.}

\label{tab:fuel_prop}
\end{table}

\begin{figure}[h!]   
    \centering
    \includegraphics[width=\textwidth,height=\textheight,keepaspectratio]{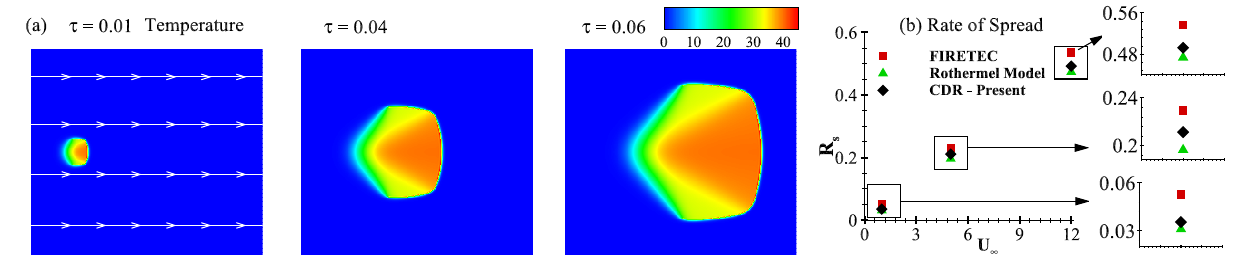}
    \caption{Evolution of the fireline under a unidirectional wind velocity field over a uniformly distributed fuel bed, evaluated across a range of \( \Phi \in (1.113\times 10^{-3},\, 2.221\times 10^{-4},\, 9.278\times 10^{-5}) \), corresponding to three different wind velocities, with \textit{Da} = 1112.735 and \( \epsilon = 0.03 \). (a) Temperature fields at selected time instants, with streamlines superimposed at \( \tau = 0.01 \), illustrate the evolving fireline structure for $\Phi = 9.278\times 10^{-5}$. (b) Comparison of three wildfire modeling approaches for estimating the rate of spread ($R_s: m/s$): FIRETEC and the Rothermel model---obtained in~\cite{pimont2012coupled}---are evaluated against predictions of the present study's CDR wildfire model, under varying wind velocities ($U_\infty: m/s$) and the initial fireline length of 50 m.}

    \label{fig:method}
\end{figure}

The simulation is conducted over a square domain of side length 320, using the following parameters: \( \epsilon = 3 \times 10^{-2} \), \( q = 1 \), \( \alpha = 10^{-3} \), \(\overline{\kappa} = 0.1\), \( \overline{T}_{\text{pc}} = 1, l_0 = 1 \), initial firefront temperature \( \overline{T} = 31 \), and wind speed \( |\overrightarrow{\mathbf{w}}| = 1 \) oriented along horizontal direction. Additionally, we consider \( C_\mu = 0.4 \), $Da = 1112.735$, and \( \Phi \in (1.113 \times 10^{-3},\, 2.221 \times 10^{-4},\, 9.278 \times 10^{-5}) \), corresponding to the three wind velocity cases in the FIRETEC simulations. The firefront is initialized as a rectangular strip of length 50 and width 2.5, centered at \( (40, 160) \), with a uniform fuel loading of 1.0 across the domain. Simulations are performed using a time step of \( \Delta \tau = 10^{-7} \) on a \( 256 \times 256 \) grid.

The results, shown in Fig.~\ref{fig:method}a, depict the spatio-temporal temperature evolution of the fireline at selected time instants, while Fig.~\ref{fig:method}b presents the quantitative comparison of the rate of spread ($R_s$) across the three wildfire modeling approaches. The CDR model's predictions are compared against data from~\cite{pimont2012coupled}, highlighting both agreements and some discrepancies. The deviations were relatively minor (particularly at lower velocities) and can be attributed to several modeling assumptions, including the use of localized artificial diffusion, spatial parameter averaging, and the omission of complex flame–wind coupling effects. Moreover, such discrepancies are expected due to the inherent limitations and simplifications of the CDR framework.~\citet{prieto2015sensitivity} presented a two-dimensional simplified CDR wildfire model and demonstrated that the governing spatially averaged physical variables can be optimized through a constrained minimization approach using data from either three-dimensional CFD simulations or experimental measurements, thereby enabling the two-dimensional model to approach more realistic results.
}
\section{Notations}
\label{appendix:notation}
\setlength{\tabcolsep}{3em}
\begin{longtable}{@{}c|c@{}}
\toprule
Notation                      & Definition                                       \\ \midrule
\endfirsthead

\multicolumn{2}{c}%
{{\tablename\ \thetable{} -- continued from previous page}} \\
\toprule
Notation                      & Definition                                       \\ \midrule
\endhead

\midrule
\multicolumn{2}{r}{{Continued on next page}} \\
\endfoot

\endlastfoot

$A$                           & Pre-exponential factor                           \\ \midrule
$E_A$                         & Activation energy                                \\ \midrule
$R$                           & Universal gas constant                           \\ \midrule
$T$                           & Absolute temperature                             \\ \midrule
$r$                           & Reaction rate                                    \\ \midrule
$Y$                           & Mass fraction of fuel                            \\ \midrule
$\rho$                        & Density                                          \\ \midrule
$\overrightarrow{\mathbf{v}}$ & Velocity vector                                  \\ \midrule
$U_\infty$                    & Maximum freestream velocity magnitude            \\ \midrule
$T_\infty$                    & Absolute ambient temperature                     \\ \midrule
$\sigma$                      & Stefan-Boltzmann constant                        \\ \midrule
\( \delta \)                  & Optical path length for radiation                \\ \midrule
$h$                           & Natural convection coefficient                   \\ \midrule
$H$                           & Heat of combustion                               \\ \midrule
$k$                           & Thermal conductivity                             \\ \midrule
$C$                           & Specific heat                                    \\ \midrule
$q$                           & Non-dimensional reaction heat                    \\ \midrule
$t_0$                         & Characteristic temporal scale                    \\ \midrule
$t_d$                         & Characteristic diffusion time scale              \\ \midrule
$t_r$                         & Characteristic fuel reaction time scale          \\ \midrule
$t_f$                         & Characteristic flow advection time scale         \\ \midrule
$l_0$                         & Characteristic spatial scale                     \\ \midrule
$\overrightarrow{\mathbf{w}}$ & Normalized velocity vector                       \\ \midrule
\( \epsilon \)             & Inverse activation energy  at $T_\infty$            \\ \midrule
$Y_0$                         & Initial mass fraction of fuel                    \\ \midrule
$\beta$                       & Normalized mass fraction of fuel                 \\ \midrule
$\beta_{max}$                 & Initial normalized mass fraction of fuel          \\ \midrule
$\hat{n}$                     & Unit normal vector                               \\ \midrule
$T_{\mathrm{pc}}$             & Phase change temperature                         \\ \midrule
$s(T)^{+}$                    & Phase change function                            \\ \midrule
$\alpha$                      & Non-dimensional natural convection coefficient   \\ \midrule
$\overline{\kappa}$           & Inverse non-dimensional conductivity coefficient \\ \midrule
$\overline{T}$                & Non-dimensional absolute temperature             \\ \midrule
$\overline{T}_{max}$          & Maximum non-dimensional absolute temperature     \\ \midrule
$\overline{T}_{\mathrm{pc}}$  & Non-dimensional phase change temperature         \\ \midrule
$s(\overline{T})^{+}$         & Non-dimensional phase change function            \\ \midrule
\textit{Da}                   & Damk\"{o}hler number                             \\ \midrule
\textit{Pe}                   & Peclet number                                    \\ \midrule
$\Phi$                        & Ratio of Damk\"{o}hler number to Peclet number   \\ \midrule
$h_x$                         & Spatial grid spacing                              \\ \midrule
$\widehat{T}$                 & Non-stiff $\overline{T}$                         \\ \midrule
$\tau$                        & Dimensionless evolution time                     \\ \midrule
$\tau^n$                      & Current temporal solution                        \\ \midrule
$\tau^{n+1}$                  & Future temporal solution                         \\ \midrule
$\Delta \tau$                 & Temporal resolution                              \\ \midrule
$\Delta^2_T$                  & Localized artificial diffusion                   \\ \midrule
$\mu$                         & Localized artificial diffusion coefficient       \\ \midrule
$C_\mu$                       & Localized artificial diffusion magnitude         \\ \midrule
$\mathbf{H_F}$                & Heaviside function for locating discontinuities  \\  \midrule 
$\widehat{\mathbf{C}}$                  & Cauchy-Green Strain tensor                       \\ \midrule
$\widehat{t}_0$                         & Initial time for computing FTLE integration     \\ \midrule
$t_1$                         & FTLE integration time period                    \\ \midrule
$\widehat{\sigma}$            & FTLE field                                      \\ \midrule
$\overline{h}_x$              & Firefront thickness                             \\ \midrule
x, y                          & Spatial coordinates                             \\ \midrule
$\xi, \eta$                   & Normalized spatial coordinates                   \\ \midrule
u, v                          & Wind velocity components along $\xi$ and $\eta$ axis  \\ \midrule
$F_T^{Y}$                     & Top firefront advecting along $-\eta$ direction     \\ \midrule
$F_B^{Y}$                     & Bottom firefront advecting along $+\eta$ direction  \\ \midrule
$F_R^{X}$                     & Right firefront advecting along $+\xi$ direction   \\ \midrule
$F_L^{X}$                     & Left firefront advecting along $-\xi$ direction     \\\midrule
$V_T^{Y}$                     & Time-averaged group velocity of top firefront     \\ \midrule
$V_B^{Y}$                     & Time-averaged group velocity of bottom firefront  \\ \midrule
$V_R^{X}$                     & Time-averaged group velocity of right firefront   \\ \midrule
$V_L^{X}$                     & Time-averaged group velocity of left firefront    \\ \midrule
$A_m$                         & Velocity magnitude                                \\ \midrule
$\lambda$                     & Wind oscillation amplitude                        \\ \midrule
$\Omega$                      & Wind oscillation frequency                        \\ \midrule
$\overline{t}_r$              & Total fuel consumption time scale                 \\ \midrule
\textit{St}                  & Strouhal number                                   \\ \midrule
$t_{left}$                  & Time taken by $F_L^{X}$ to reach the left  boundary   \\ \midrule
$t_{right}$                 & Time taken by $F_R^{X}$ to reach the right boundary   \\ \midrule
$t_{top}$                   & Time taken by $F_T^{Y}$ to reach the top boundary     \\ \midrule
$TF$                        & Transfer function                                   \\ \midrule
$\overline{R}, \phi$        & Real and phase angle of TF                          \\ \midrule  
$\overline{R}_R, \phi_R$    & Real and phase angle of TF for a right firefront    \\ \midrule
$R_s$                        & Rate of Spread                                     \\ \midrule
$R_0$                        & Rate of Spread at zero-slope and no wind condition   
\\\bottomrule
\caption{Nomenclature of parameters used in the wildfire combustion model.}
\label{nomen} 
\end{longtable}

%\bibliographystyle{jfm}
\bibliography{jfm-instructions}

\end{document}